\documentclass{article} 
\usepackage{iclr2026_conference,times}

\usepackage{amsmath,amsfonts,bm}

\def\eqref#1{equation~\ref{#1}}

\def\1{\bm{1}}

\DeclareMathAlphabet{\mathsfit}{\encodingdefault}{\sfdefault}{m}{sl}
\SetMathAlphabet{\mathsfit}{bold}{\encodingdefault}{\sfdefault}{bx}{n}

\usepackage{hyperref}
\usepackage{url}
\usepackage{enumitem}
\usepackage[utf8]{inputenc} 
\usepackage[T1]{fontenc}    
\usepackage{hyperref}       
\usepackage{url}            
\usepackage{booktabs}       
\usepackage{amsfonts}       
\usepackage{nicefrac}       
\usepackage{microtype}      
\usepackage{xcolor}         
\usepackage{enumitem}
\usepackage{amsmath, amssymb} 
\usepackage{bm} 
\usepackage{graphicx} 
\usepackage{multirow}
\usepackage{colortbl}
\usepackage{hhline}
\usepackage{longtable}
\usepackage{caption}
\usepackage{algorithm}
\usepackage{algpseudocode}
\usepackage{graphicx} 
\usepackage{booktabs} 
\usepackage{array}    
\usepackage{xcolor}   
\usepackage{caption}  
\usepackage{pgfplots}
\usepackage{makecell} %
\usepackage{amsfonts}       
\usepackage{adjustbox}      
\usepackage{siunitx}
\usepackage{wrapfig}
\usepackage{amsthm} 
\usepackage{fancyhdr}
\usepackage{pifont}
\usepackage{xspace}
\usepackage{titletoc}
\usepackage{tcolorbox}
\usepackage{tablefootnote}\usepackage{fontawesome}   
\usepackage{tikz}
\usepackage{fontawesome}
\usepackage{graphicx}
\usepackage{subcaption}
\usepackage{longtable}
\usepackage{booktabs}
\usepackage{tabularx}
\pgfplotsset{compat=1.17}
\newcolumntype{C}[1]{>{\centering\arraybackslash}p{#1}}

\usepackage{tikz}

\definecolor{mygray}{gray}{.92}
\definecolor{ForestGreen}{RGB}{34,139,34}
\definecolor{Forestred}{RGB}{220,50,50}

\newcommand{\fg}[1]{\mathbf{\mathcolor{ForestGreen}{#1}}}

\definecolor{darkpink}{RGB}{255, 20, 147}

\newcommand{\benchname}{\textsc{AudioMarathon}}
\title{AudioMarathon: A Comprehensive Benchmark for Long-Context Audio Understanding and Efficiency in Audio LLMs}

\author{
    Peize He$^{1}$\footnotemark[1]\quad 
    Zichen Wen$^{1,2}$\footnotemark[1]\quad 
    Yubo Wang$^{1}$\footnotemark[1]\quad
    Yuxuan Wang$^{1}$\quad
    Xiaoqian Liu$^{1,3}$ \\
    \textbf{Jiajie Huang}$^{1}$ \quad
    \textbf{Zehui Lei}$^{1}$\quad
    \textbf{Zhuangcheng Gu}$^{4}$\quad
    \textbf{Xiangqi Jin}$^{1}$\quad
    \textbf{Jiabing Yang}$^{5}$ \\
    \textbf{Kai Li}$^{6}$\quad
    \textbf{Zhifei Liu}$^{1}$\quad
    \textbf{Weijia Li}$^{7,2}$\quad
    \textbf{Cunxiang Wang}$^{6}$\quad 
    \textbf{Conghui He}$^{2}$\quad 
    \textbf{Linfeng Zhang}$^{1}$\footnotemark[2]
    \vspace{4pt} \\
    $^{1}$Shanghai Jiao Tong University \quad 
    $^{2}$Shanghai AI Laboratory \quad 
    $^{3}$Northeastern University 
    \\
    $^{4}$Carnegie Mellon University\quad
    $^{5}$University of Chinese Academy of Sciences \\
    $^{6}$Tsinghua University \quad
    $^{7}$Sun Yat-sen University \quad
}

\iclrfinalcopy 
\begin{document}

\maketitle

{
\renewcommand{\thefootnote}{\fnsymbol{footnote}}
  \footnotetext[1]{Equal Contribution.} 
  \footnotetext[2]{Corresponding author: zhanglinfeng@sjtu.edu.cn}
}

\begin{abstract}
Processing long-form audio is a major challenge for Large Audio Language models (LALMs). These models struggle with the quadratic cost of attention ($\mathcal{O}(N^2)$) and with modeling long-range temporal dependencies. Existing audio benchmarks are built mostly from short clips and do not evaluate models in realistic long context settings. To address this gap, we introduce \textbf{\benchname{}}, a benchmark designed to evaluate both \emph{understanding} and \emph{inference efficiency} on long-form audio. \benchname{} provides a diverse set of tasks built upon three pillars: long-context audio inputs with durations ranging from 90.0 to 300.0 seconds, which correspond to encoded sequences of 2,250 to 7,500 audio tokens, respectively, full domain coverage across speech, sound, and music, and complex reasoning that requires multi-hop inference. 
We evaluate state-of-the-art LALMs and observe clear performance drops as audio length grows. 
We also study acceleration techniques and analyze the trade-offs of token pruning and KV cache eviction. The results show large gaps across current LALMs and highlight the need for better temporal reasoning and memory-efficient architectures. 
We believe \benchname{} will drive the audio and multimodal research community to develop more
advanced audio understanding models capable of solving complex audio tasks. 

\raisebox{-0.3\height}{\includegraphics[width=0.4cm]{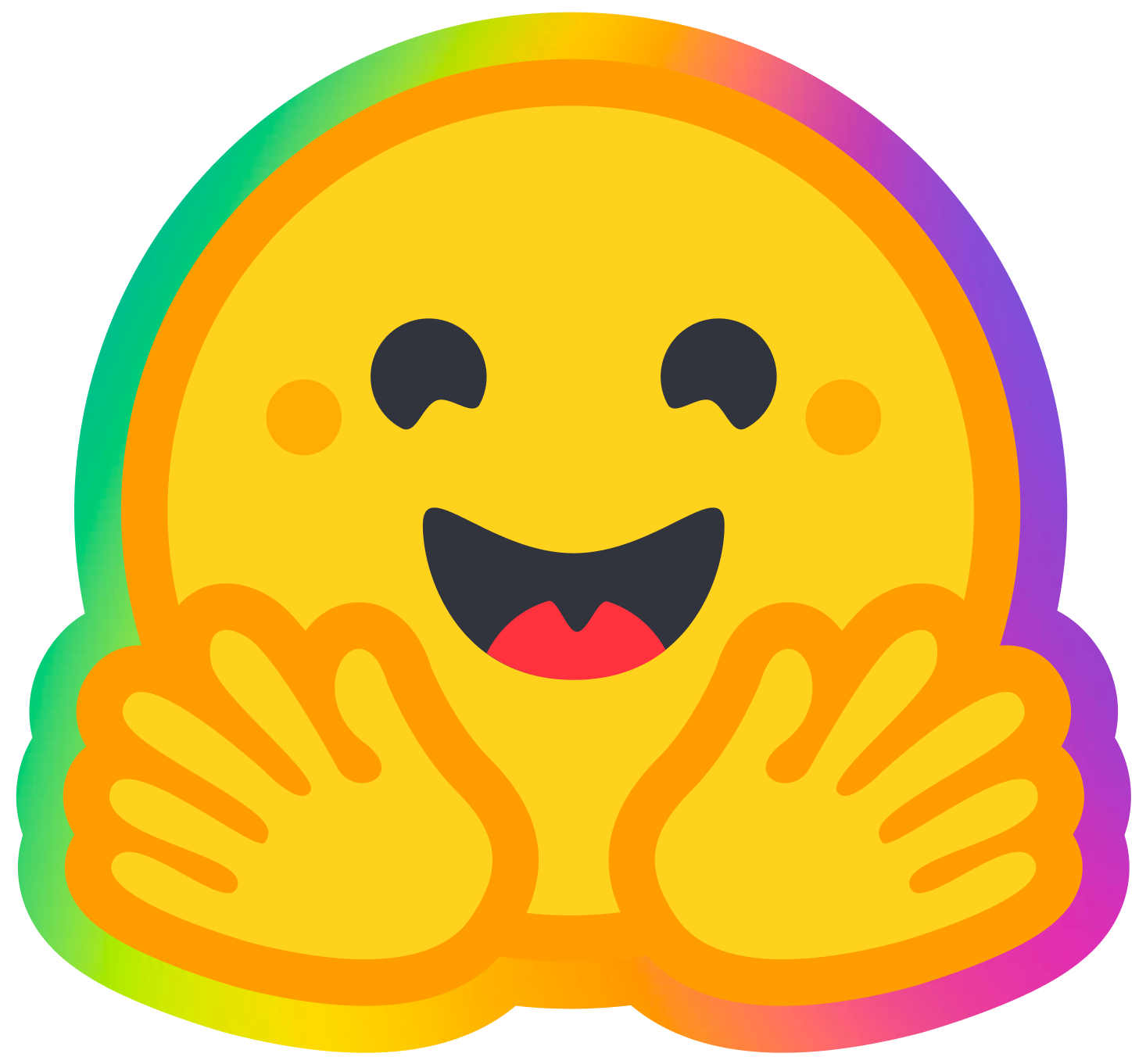}} \small \textbf{\mbox{Data \& Dataset Card:}} \href{https://huggingface.co/datasets/Hezep/AudioMarathon}{huggingface.co/datasets/Hezep/AudioMarathon} \\
\vspace{1em}
\raisebox{-0.3\height}{\hspace{-0.11cm}\includegraphics[width=0.45cm]{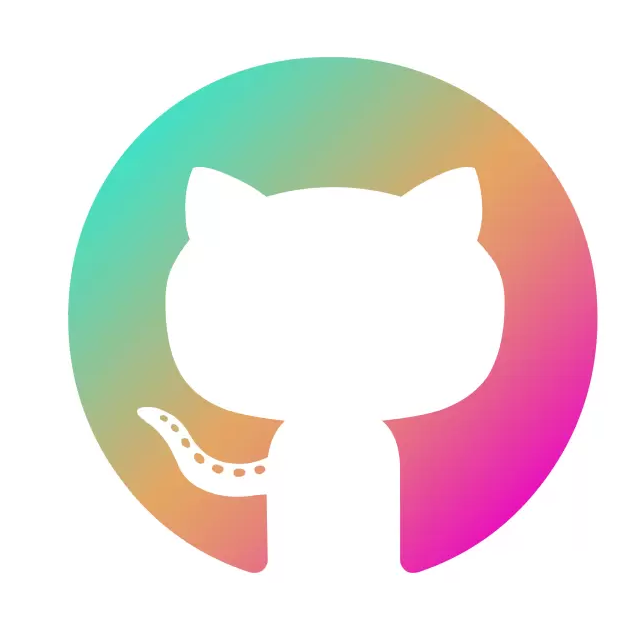}} \small \textbf{\mbox{Code Repository:}} \href{https://github.com/DabDans/AudioMarathon}{github.com/DabDans/AudioMarathon}

\end{abstract}

\section{Introduction}
\label{sec:introduction}
Multimodal Large Language Models (MLLMs), benefiting from the powerful understanding and reasoning abilities of large language models, have demonstrated remarkable capabilities in understanding and processing various data modalities~\citep{alayrac2022flamingo, li2023blip, liu2023visual, chen2024mj, kang2025legion, zhang2024ocr, wen2024aidbench, li2025tactic}. With audio being a key area of advancement, the ability to comprehend spoken language, environmental sounds, and music has opened up new frontiers for applications ranging from advanced speech recognition~\citep{radford2023robust} to sophisticated audio-based reasoning~\citep{borsos2023audiolm}. 

However, a significant and persistent challenge remains: the effective processing of long-form audio inputs. As the duration of audio increases, Large Audio Language Models (LALMs) face a dual challenge of escalating computational and memory costs~\citep{vaswani2017attention}, coupled with the inherent difficulty of capturing and modeling extended temporal dependencies~\citep{beltagy2020longformer, zaheer2020big}. This bottleneck severely limits their practical application in real-world scenarios such as analyzing meetings, podcasts, or extended dialogues.
\begin{figure*}[!t]
    \centering
    \includegraphics[width=\linewidth]{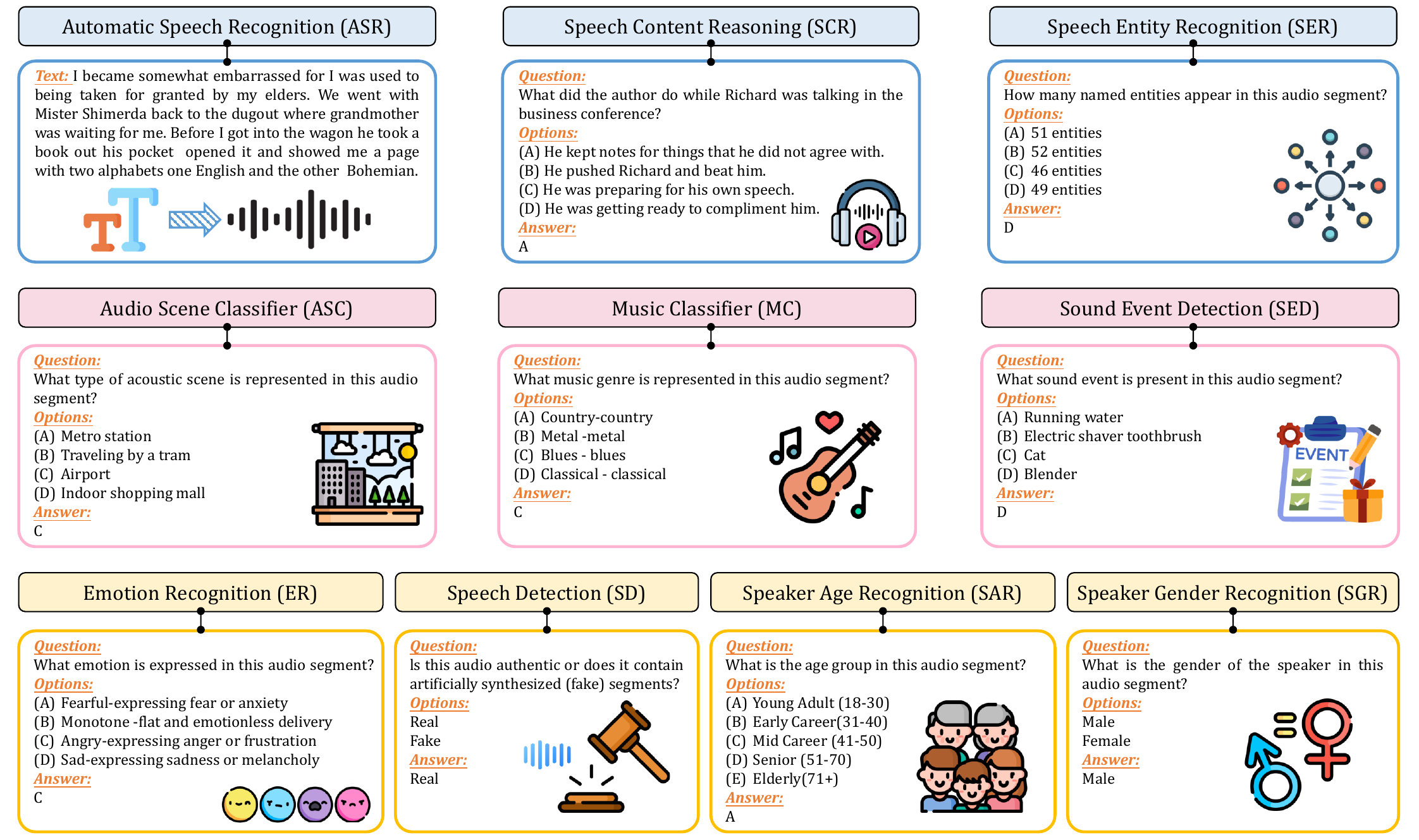} 
    \vspace{-1.5em}
    \caption{Overview of the \benchname{}. \benchname{} extends short audio clips to long-form audio with a diverse range of task categories, offering a comprehensive and practical assessment of audio intelligence in real-world scenarios.}
    \label{fig:BenchQA}
    \vspace{-1.95em}
\end{figure*}
A major factor hindering progress in this domain is the lack of comprehensive benchmarks designed to evaluate the long audio capabilities of LALMs rigorously. Existing audio benchmarks predominantly consist of short clips, typically only a few seconds long~\citep{weck2024muchomusic,sakshi2024mmau,yang2024air,wang2024audiobench}. While valuable, these benchmarks fail to assess a model's ability to maintain coherence, reason over long time spans, and manage computational resources efficiently when faced with minute-scale and even hour-scale audio inputs. This gap leaves a critical aspect of model performance unevaluated and obstructs the development of more robust and scalable audio understanding models. 

To address this critical gap, we introduce \textbf{\benchname{}}, a comprehensive audio benchmark meticulously designed to evaluate LALMs on long-context audio understanding and inference efficiency. \benchname{} is built on three foundational pillars: \textbf{\ding{182} Long-form Audio Context}, featuring audio durations ranging from 90.0 to 300.0 seconds to simulate realistic scenarios; \textbf{\ding{183} Full Domain Coverage}, encompassing a diverse range of audio types including speech, environmental sounds, and music, as well as comprehensive task coverage spanning ten representative sub-tasks (ASR, SCR, SER, MC, ASC, SED, ER, SD, SAR, SGR) across Speech Context Understanding, Audio Scene Understanding, and Voice Characteristic Identification; and \textbf{\ding{184} Complex Reasoning}, incorporating multi-hop inference tasks that require models to connect disparate pieces of information across extended temporal windows.

Beyond just establishing a challenging new benchmark, this work also investigates crucial aspects of inference efficiency for long audio. We systematically evaluate a suite of state-of-the-art Audio LLMs~\citep{chu2023qwen,abouelenin2025phi,xu2025qwen2}, analyzing their performance degradation as input length increases. 
Furthermore, we explore and quantify the effectiveness and trade-offs of various cost-reduction strategies, including inference-time \textbf{Token pruning}~\citep{chen2024image,zhang2024sparsevlm,wen2025stop,wen2025efficient} and \textbf{KV-cache eviction} techniques~\citep{li2024snapkv}. 
Our findings reveal substantial performance gaps among current models in long-context scenarios and underscore the pressing need for improved temporal reasoning and memory-efficient processing.

By providing a unified and challenging evaluation suite, we aim to catalyze future research. We release \benchname{} to the community to foster the development of the next generation of scalable, efficient, and robust LALMs capable of truly understanding the rich, continuous tapestry of the auditory world. Our main contributions are summarized as follows:
\begin{itemize}[leftmargin=10pt, topsep=0pt, itemsep=1pt, partopsep=1pt, parsep=1pt]
    \item \benchname{} is presented as a comprehensive benchmark for long audio understanding, characterized by extended audio durations, diverse domain coverage, and complex reasoning tasks.
    \item Our work thoroughly evaluates state-of-the-art LALMs on \benchname{}, revealing the specific challenges encountered when processing long audio inputs.
    \item In addition, we systematically analyze various inference efficiency techniques, such as token pruning and KV-cache eviction, to quantify their effectiveness and trade-offs.
\end{itemize}

\section{\benchname}\label{sec:method}
\begin{figure*}[!t]
    \centering
    \includegraphics[width=\linewidth]{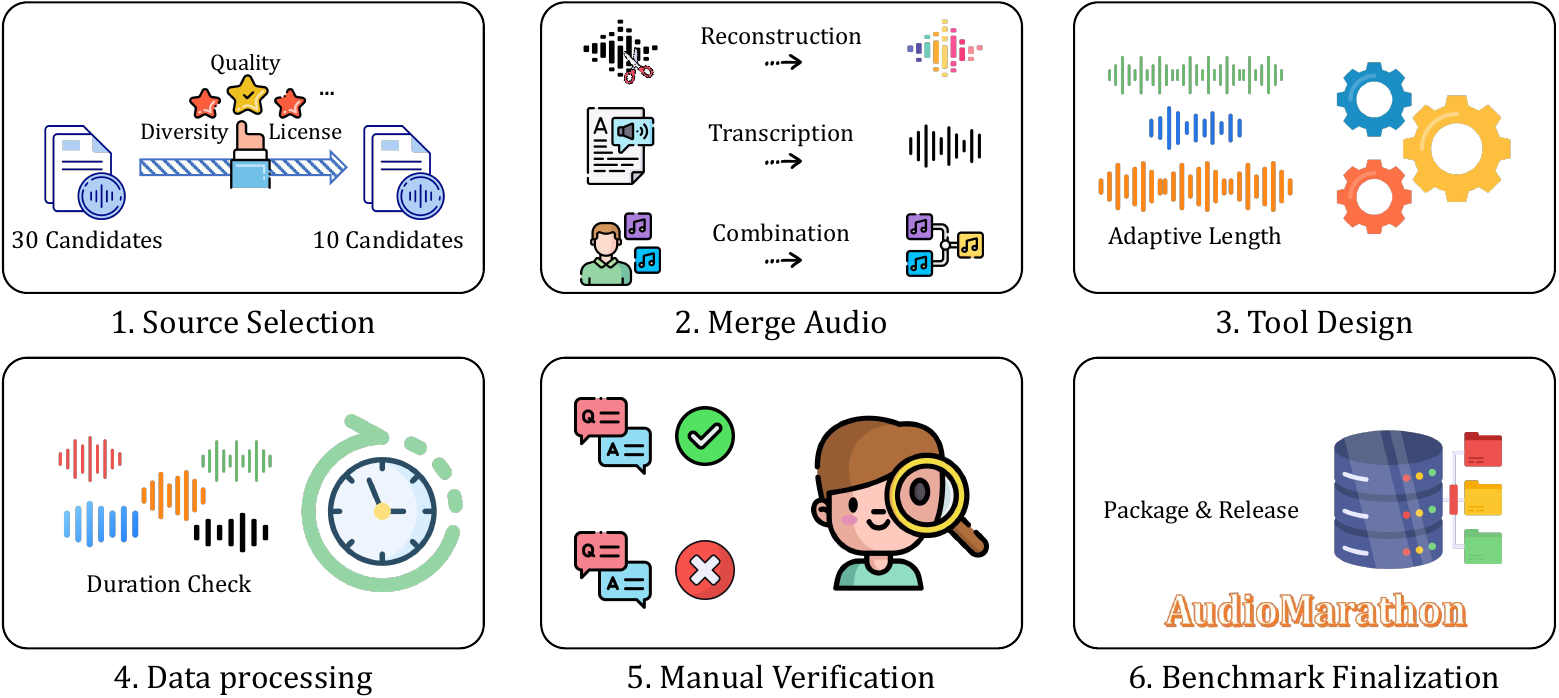} 
    \vspace{-4mm}
    \caption{The six-stage data pipeline for constructing the \benchname{}}
    \label{fig:pipeline}
    \vspace{-4mm}
\end{figure*}

\begin{figure}[htbp] 
    \centering
    \begin{subfigure}[b]{0.48\textwidth}
        \centering
        \includegraphics[width=\linewidth]{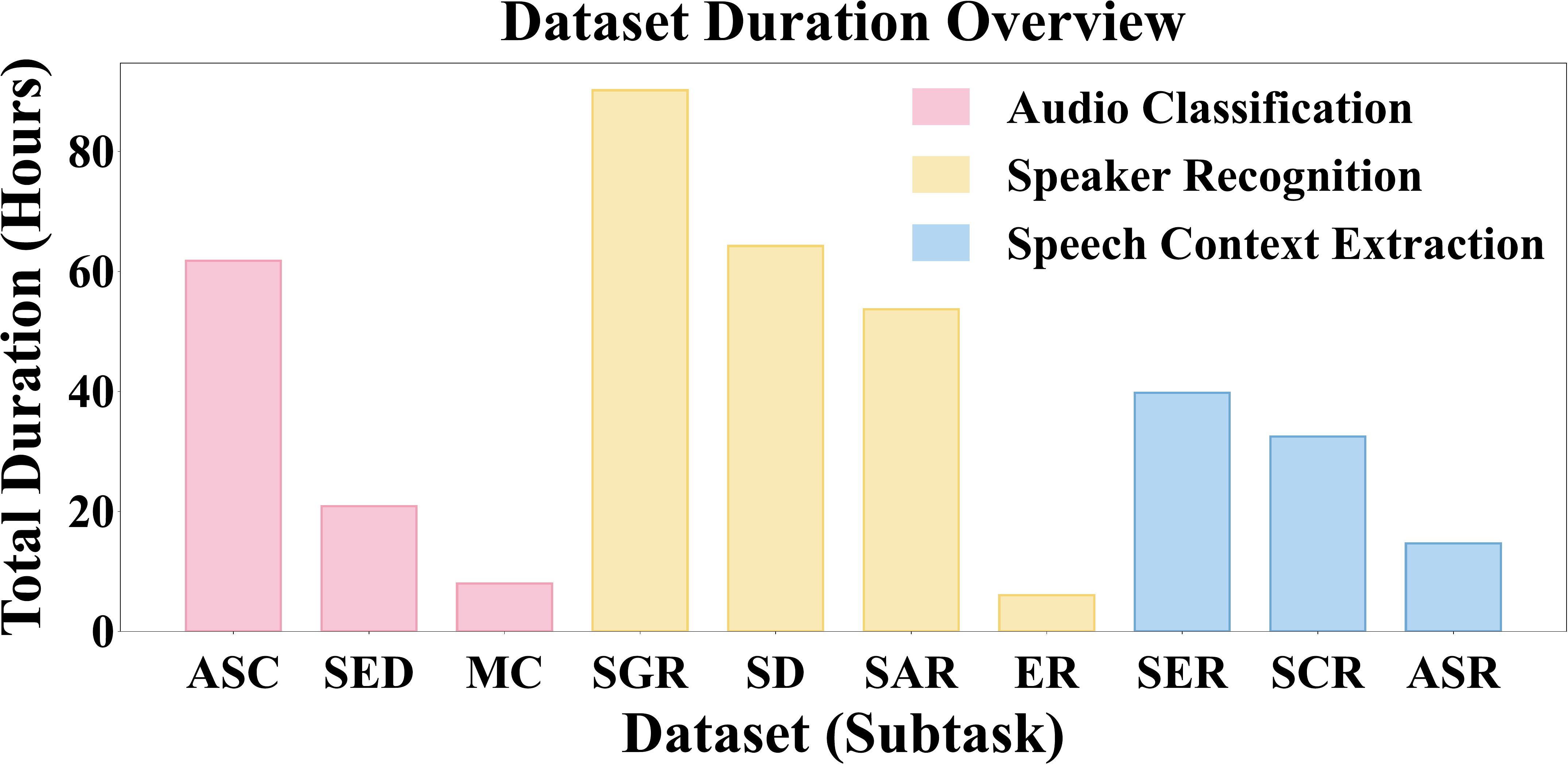} 
        \vspace{-5mm}
        \caption{Per dataset duration in \benchname{}}
        \label{fig:subfig1}
    \end{subfigure}
    \hfill 
    \begin{subfigure}[b]{0.48\textwidth}
        \centering
        \includegraphics[width=\linewidth]{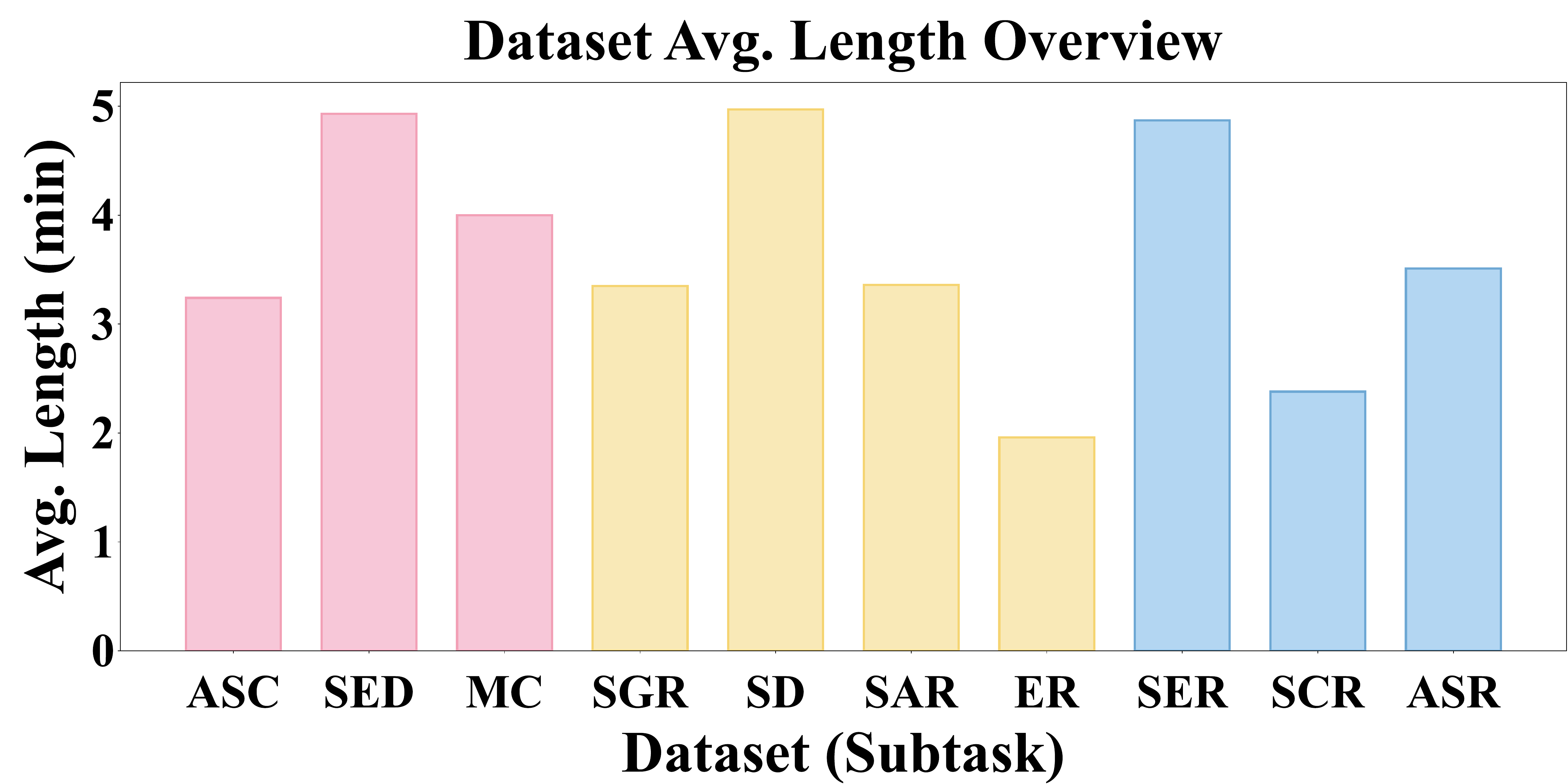} 
        \vspace{-5mm}
        \caption{Per dataset average length in \benchname{}}
        \label{fig:subfig2}
    \end{subfigure}
    \vspace{-2mm}
    \caption{Per dataset duration and average length in \benchname{}}
    \label{fig:side_by_side}
    \vspace{-3mm}
\end{figure}

\begin{wrapfigure}{r}{0.41\linewidth}
    \vspace{-3.5em}  
    \centering
    \includegraphics[width=\linewidth]{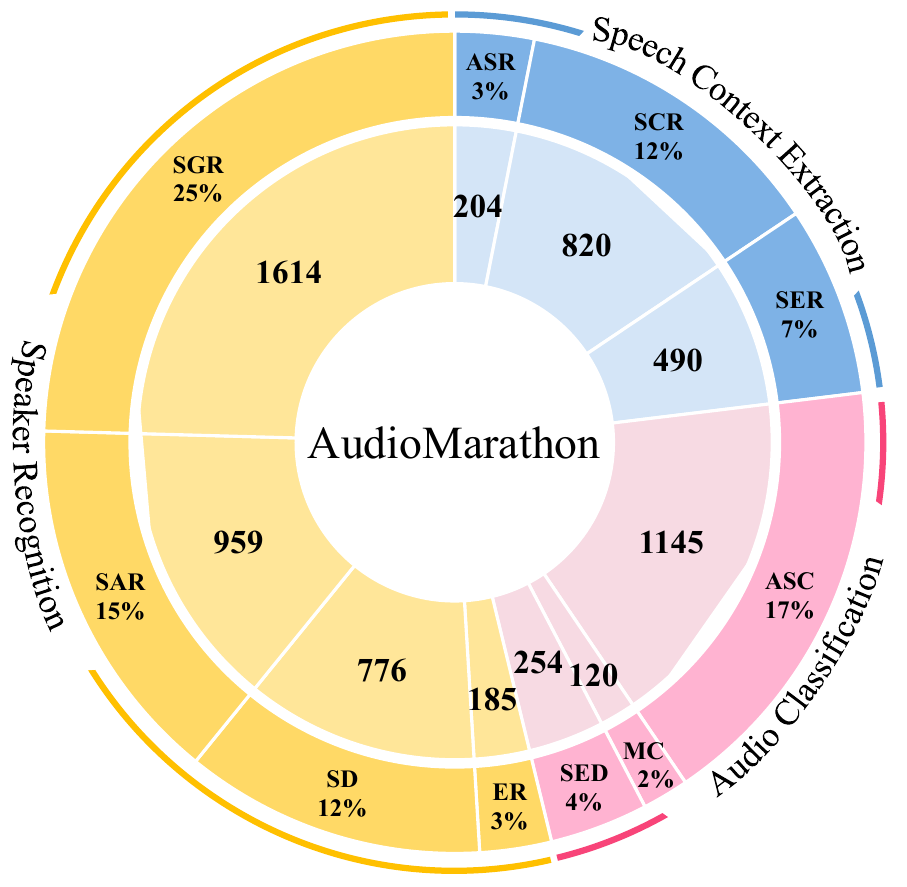}
    \vspace{-6.5mm}
    \caption{Task composition of \benchname{} by category}
    \label{fig:LAB}
    \vspace{-2em}
\end{wrapfigure}
\subsection{Overview}
Existing audio benchmarks predominantly comprise short audio clips, often only a few seconds, thereby failing to capture the complexity of real-world scenarios such as meetings, podcasts, and extended dialogues.
To close the pronounced gap in benchmarks for long-form audio understanding, we present \textbf{\benchname{}}, a comprehensive suite designed to evaluate the advanced capabilities of LALMs. 
The construction of \benchname{} follows a rigorous six-stage pipeline (Figure \ref{fig:pipeline}), ensuring diversity, difficulty, and high annotation quality. Figure \ref{fig:LAB} and Table \ref{tab:mmau_stats} summarize the final composition across task categories, while Table \ref{tab:my_awesome_table} and Table \ref{tab:compare} compare \benchname{} against existing benchmarks.

\subsection{Data Collection and Annotation}
We adopt a rigorous multi-stage framework to construct \textbf{\benchname{}}, detailed below.
\noindent \textbf{Step 1. Source Selection.} From 30 candidate datasets, we selected ten subsets according to task coverage and acoustic diversity. The tasks are grouped into three categories: Speech Context Understanding, Audio Scene Understanding, and Voice Characteristic Identification, ensuring both practical relevance and expert-level reasoning challenges.
\\
\noindent \textbf{Step 2. Merge Audio.} Considering the characteristics of different tasks, we designed specific concatenation logic to merge individual clips into longer sequences.  
\\
\noindent \textbf{Step 3. Tool Design.} We developed a custom concatenation script to automate merging process. The tool flexibly supports generating sequences of variable length within the constraints of the source material, allowing to adapt sequence duration for different experimental settings.  
\\
\noindent \textbf{Step 4. Data Processing.} Each audio file was paired with a task-specific prompt and multiple-choice options. Option generation followed customized strategies tailored to each task, and the implementation has been released as open source. Model predictions were evaluated by two criteria: (i) exact match to a provided choice, or (ii) inclusion of the complete correct option without any extraneous information.  
\\
\noindent \textbf{Step 5. Manual Verification.} To ensure data quality, 10\% (at least 20) samples per sub-dataset were randomly reviewed using the criteria detailed in Appendix~\ref{sec:dataset_details}. Any dataset failing inspection was reconstructed and revalidated until all checked samples passed.  
\\
\noindent \textbf{Step 6. Benchmark Finalization.} From the fully annotated QA pairs, 6,567 instances were selected to ensure balanced coverage of all 10 tasks and audio types. The concatenated files had durations from 90.0 to 300.0 seconds, balancing long-context evaluation with computational feasibility.

\subsection{Comparison with Other Benchmarks}
\begin{table*}[!h]
    \caption{Comparison of audio datasets in terms of duration, size, average audio length, and domain coverage (speech, sound, and music).}
    \vspace{-2mm}
    \label{tab:my_awesome_table}
    \centering
     \scriptsize
    \renewcommand{\arraystretch}{1}
    \setlength{\tabcolsep}{4pt}
    \begin{tabular*}{\linewidth}{@{\extracolsep{\fill}} lcccccc}
    \toprule
    \multirow{2}{*}{Tasks} & 
    \multirow{2}{*}{Duration} & 
    \multirow{2}{*}{Size} & 
    \multicolumn{3}{c}{Domain} & 
    \multirow{2}{*}{\shortstack{Average audio \\ duration}} \\
    \cmidrule(lr){4-6}
    & & & Speech & Sound & Music & \\
    \midrule
    MuChoMusic~\citep{weck2024muchomusic}   & 5.1h   & 1.1k  & $\times$ & $\times$ & \checkmark & 25.7 sec \\
    BLAB~\citep{ahia2025blab}         & 833h   & 1.6k  & \checkmark & $\times$ & $\times$ & 51.0 min \\
    MMAR~\citep{ma2025mmar}           & 5.5h   & 1k    & \checkmark & \checkmark & \checkmark & 19.4 sec \\
    MMSU~\citep{wang2025mmsu}         & 9.73h  & 5k    & \checkmark & $\times$ & $\times$ & 7.0 sec \\
    MMAU~\citep{sakshi2024mmau}       & 28.16h & 10k   & \checkmark & \checkmark & \checkmark & 10.1 sec \\
    AIR-Bench~\citep{yang2024air}     & 251.6h & 21k   & \checkmark & \checkmark & \checkmark & 35.2 sec \\
    AudioBench~\citep{wang2024audiobench} & 400h & 100k & \checkmark & \checkmark & \checkmark & 14.0 sec \\
    \textbf{AudioMarathon}\textit{ (ours)} 
                     & 392h   & 6.6k  & \checkmark & \checkmark & \checkmark & 212.8 sec \\
    \bottomrule
    \vspace{-6mm}
  \end{tabular*}
\end{table*}
\noindent \textbf{Long Audio Understanding.} Public audio benchmarks mostly use second-level clips (e.g., MMAR 19.4 s, MMAU 10.1 s, MMSU 7.01 s, AudioBench 14 s), which miss minute-scale complexity. BLAB includes long audio ($\sim$51.0 min) but is speech-centric. \benchname{} targets realistic long-form use with durations ranging from 90.0 to 300.0 seconds and supports flexible duration control.

\noindent \textbf{Full Domain Coverage.} Audio spans three domains: speech, sound, and music. Most benchmarks cover one or two, limiting cross-domain robustness. Our proposed \benchname{} covers all three with balanced sampling for comprehensive evaluation and cross-domain studies.

\noindent \textbf{Multi-Hop Inference.} We include an audio version of RACE generated via Text-to-Speech (Kokoro-82M~\citep{nayak2025kokoro}), preserving RACE's multi-hop reasoning while adding long-term acoustic dependencies—a stricter test of comprehension, memory, and reasoning.
\begin{table*}[!h]
\caption{Comparison of \benchname{} with existing audio understanding and reasoning benchmarks across key properties and capabilities.}
\vspace{-2mm}
\resizebox{\linewidth}{!}{
\begin{tabular}{lcccccccc}
\toprule
\textbf{Capability}                                     & \textbf{\benchname{}} & \textbf{MuChoMusic} & \textbf{BLAB} & \textbf{MMAR} & \textbf{MMSU} & \textbf{MMAU} & \textbf{AIR-Bench} & \textbf{AudioBench} \\ \midrule
Long Audio Understanding
& \textcolor{green}{\checkmark}    & \textcolor{red}{$\times$}         & \textcolor{green}{\checkmark}    & \textcolor{red}{$\times$}         & \textcolor{red}{$\times$}        & \textcolor{red}{$\times$}         & \textcolor{red}{$\times$}        & \textcolor{red}{$\times$}     
\\
Full Domain Coverage                      
& \textcolor{green}{\checkmark}    & \textcolor{red}{$\times$}          & \textcolor{red}{$\times$}        & \textcolor{green}{\checkmark}      & \textcolor{green}{\checkmark}    & \textcolor{green}{\checkmark}      & \textcolor{green}{\checkmark}    & \textcolor{red}{$\times$}          
\\
Multi-Hop Inference   
& \textcolor{green}{\checkmark}    & \textcolor{red}{$\times$}          & \textcolor{red}{$\times$}        & \textcolor{red}{$\times$}         & \textcolor{red}{$\times$}        & \textcolor{red}{$\times$}          & \textcolor{red}{$\times$}        & \textcolor{red}{$\times$}
\\
Speaker attribute coverage
& \textcolor{green}{\checkmark}    & \textcolor{red}{$\times$}          & \textcolor{red}{$\times$}        & \textcolor{red}{$\times$}         & \textcolor{green}{\checkmark}         & \textcolor{red}{$\times$}          & \textcolor{green}{\checkmark}        & \textcolor{green}{\checkmark}
\\
Contain deepfake audio
& \textcolor{green}{\checkmark}    & \textcolor{red}{$\times$}          & \textcolor{red}{$\times$}        & \textcolor{red}{$\times$}         & \textcolor{red}{$\times$}        & \textcolor{red}{$\times$}          & \textcolor{green}{\checkmark}        & \textcolor{red}{$\times$}
\\
Complex task hierarchy
& \textcolor{green}{\checkmark}    & \textcolor{red}{$\times$}          & \textcolor{red}{$\times$}        & \textcolor{green}{\checkmark}         & \textcolor{green}{\checkmark}         & \textcolor{green}{\checkmark}          & \textcolor{green}{\checkmark}        & \textcolor{red}{$\times$}
\\
\makecell{Emotional and Semantic Understanding}
& \textcolor{green}{\checkmark}    & \textcolor{red}{$\times$}          & \textcolor{green}{\checkmark}        & \textcolor{green}{\checkmark}         & \textcolor{green}{\checkmark}        & \textcolor{green}{\checkmark}          & \textcolor{green}{\checkmark}        & \textcolor{green}{\checkmark}
\\ \hline
\end{tabular}}
\label{tab:compare}
\vspace{-3mm}
\end{table*}

\vspace{-3mm}
\section{Experiments and Evaluations}
\label{sec:exp}

\vspace{-2mm}
\noindent \textbf{Models.}
We compare 16 recent Large Audio Language Models (LALMs), including ten open-source models and six closed-source models. The open-source models are {Phi-4-Multimodal}~\citep{phi4}, {Qwen2.5-Omni-3B}~\citep{qwen2.5-omni}, and {Aero-1-Audio}~\citep{li2025aero}. {Phi-4-Multimodal} and {Qwen2.5-Omni-3B} are multi-modal large language models, while {Aero-1-Audio} is a compact audio language model designed for audio-centered tasks. The proprietary models are from the Gemini family: {Gemini-2.5-Pro}~\citep{comanici2025gemini}, {Gemini-2.5-Flash}~\citep{comanici2025gemini}, {Gemini-2.0-Flash}~\citep{gemini2.X}, and {GPT-4o}. All are multi-modal models, with {Gemini-2.5-Flash} and {Gemini-2.0-Flash} optimized for faster inference.

\noindent \textbf{Evaluation Metrics.}
Our evaluation considers two dimensions: task performance and inference efficiency. For task performance, we adopt standard metrics per task: F1-score for classification and MCQs, Word Accuracy Rate (WAR) for ASR, and macro F1-score for audio event detection to balance precision and recall across classes. Inference efficiency is assessed via latency and peak GPU memory usage. We also report speedup over a vanilla model.

\noindent \textbf{Evaluation Setup.}
To conduct the ASR task, evaluations are performed on \textit{test} subset of LibriSpeech-long~\citep{park2024long} after filtering.
Except for ASR, all tasks are framed as MCQs with a single correct answer. SD and SGR provide two options, SAR provides five, and all other tasks use four. 
For each instance, the model receives the full audio along with an instruction-following prompt presenting a question and four labeled options. 
The model must select one option, and to mitigate positional bias, the option order is randomized.

\begin{table*}[!t]
\centering
\caption{Performance comparison of models on AudioMarathon across tasks, grouped into Speech Content Extraction (SER, SCR, ASR), Audio Classification (SED, MC, ASC), and Speaker Information Modeling (SD, ER, SAR, SGR). The Avg. column shows the mean score across all tasks. Best scores are in \textbf{bold}, second-best are 
\underline{underlined}.}
\vspace{-2.5mm}
\centering
\renewcommand{\arraystretch}{1.05}

\newcolumntype{C}[1]{>{\centering\arraybackslash}p{#1}}
\newlength{\mycolwidthA}
\newlength{\mycolwidthB}
\settowidth{\mycolwidthA}{150.0}
\settowidth{\mycolwidthB}{150.0}
\setlength{\mycolwidthA}{1.2\mycolwidthB}

\resizebox{\linewidth}{!}{
\begin{tabular}{l *{3}{C{\mycolwidthA}} *{7}{C{\mycolwidthB}} c} 
\toprule \toprule
\multirow{2}{*}{\textbf{Models}} & \multicolumn{3}{c}{\textbf{Speech Content Extraction}} & \multicolumn{3}{c}{\textbf{Audio Classification}} & \multicolumn{4}{c}{\textbf{Speaker Information Modeling}} & \multirow{2}{*}{\textbf{Avg.}} \\
\cmidrule(lr){2-4} \cmidrule(lr){5-7} \cmidrule(lr){8-11}
& \textbf{SER} & \textbf{SCR} & \textbf{ASR} & \textbf{SED} & \textbf{MC} & \textbf{ASC} & \textbf{SD} & \textbf{ER} & \textbf{SAR} & \textbf{SGR} & \\

\midrule \midrule
\multicolumn{12}{c}{\textbf{Open-source Audio LLMs}} \\ \midrule \midrule
Phi-4-Multimodal & 18.4 & 69.3 & {92.7} & 55.1 & 46.7 & 23.4 & 26.4 & 27.3 & 26.6 & 91.1 & 47.7 \\
Qwen2.5-Omni-3B & 25.2 & 82.3 & {94.7} & \underline{70.2} & \underline{97.4} & \underline{69.3} & 67.3 & 39.6 & 29.1 & 97.2 & \underline{67.2} \\
Qwen2.5-Omni-7B  & 26.3 & \textbf{85.1} & {\textbf{98.1}} & \textbf{78.4} & \textbf{100.0} & \textbf{72.2} & \textbf{72.3} & \underline{53.4} & 21.4 & \underline{98.0} & \textbf{70.5}\\
Audio-Flamingo-2 & 26.8 & 39.8 & {1.0} & 27.1 & 66.8 & 29.7 & 45.9 & 13.1 & 20.3 & 85.1 & 35.6 \\
Audio-Flamingo-3 & 21.7 & 78.9 & {94.3} & 59.5 & {97.0} & 54.1 & 33.7 & \textbf{54.3} & \textbf{40.7} & 96.2 & 63.0 \\
Gemma-3n-E2B-it & 22.5 & 51.6 & {91.3} & 50.2 & 56.8 & 28.2 & 35.1 & 15.2 & 12.2 & 91.6 & 45.5\\
Gemma-3n-E4B-it & 19.0 & 56.9 & {93.2} & 50.2 & 71.9 & 31.7 & 35.9 & 18.9 & 21.8 & 93 & 49.3\\
Voxtral-Mini-3B-2507 & 24.3 & 71.1 & {96.8} & 71.0 & 83.8 & 27.2 & \underline{68.0} & 29.7 & 30.7 & 71.0 & 57.4 \\
Baichuan-Omni-1.5 & 12.4 & 11.2 & {86.5} & 45.7 & 52.0 & 25.8 & 49.2 & 18.9 & 10.2 & 81.5 & 39.3 \\
Aero-1-Audio & 17.9 & 56.6 & {43.7} & 55.0 & 83.9 & 39.9 & 33.7 & 32.0 & 17.8 & 47.5 & 42.8\\    
\midrule \midrule
\multicolumn{12}{c}{\textbf{Close-source Audio LLMs}} \\ \midrule \midrule
GPT-4o-Audio (Preview 2024-10-01) & 25.8 & 61.4 & {94.4} & 50.7 & 59.5 & 40.8 & 32.5 & 22.5 & 17.2 & 69.2 & 47.4 \\
GPT-4o-Audio (Preview 2024-12-17)& 25.7 & 60.2 & {94.7} & 51.2 & 67.6 & 41.9 & 30.8 & 21.8 & 19.9 & 73.1& 48.7 \\
Gemini-2.0-Flash-Lite& 23.7 & 65.6 & {\underline{97.4}} & 60.9 & 86.9 & 43.4 & 34.5 & 17.3 &19.0 & 82.1 & 53.1 \\
Gemini-2.0-Flash & \textbf{30.9} & 71.8 & {96.4} & 68.1 & 88.5 & 54.1 & 32.1 & 20.1 & \underline{39.2} & 93.1 & 59.4 \\
Gemini-2.5-Flash-Lite & \underline{30.3} & 64.0 & {96.5} & 68.0 & 64.8 & 36.8 & 33.9 & 14.6 & 19.6 & 77.9 & 50.6\\
Gemini-2.5-Flash & 28.1 & \underline{83.6} & {96.8} & 69.2 & 79.3 & 40.8 & 33.1 & 31.9 & 34.3 & \textbf{99.3} & 59.6\\

\midrule \midrule

Human Evaluation & 45.1 & 88.1 & -- & 96.2 & 100.0 & 100.0 & 100.0 & 90.8 & 71.4 & 97.0 & 87.6 \\
\bottomrule 
\end{tabular}}
\label{tab:AudioMarathon_results_categories_avg}
\vspace{-1.6em}
\end{table*}

\begin{table}[!t]
\small
\centering
\caption{Performance comparison of three open-sourced LALMs across token pruning methods and ratios on AudioMarathon tasks, grouped into Speech Content Extraction (SER, SCR, ASR), Audio Classification (SED, MC, ASC), and Speaker Recognition (SD, ER, SAR, SGR). F1-score (0-100) is the primary metric, except for ASR, where Word Accuracy Rate (WAR) is used. The Avg. column shows the mean score across available tasks. Best scores within each pruning ratio are in \textbf{bold}.}
\vspace{-0.8em}
\setlength{\tabcolsep}{5pt}
\renewcommand{\arraystretch}{0.85}

\newcolumntype{C}[1]{>{\centering\arraybackslash}p{#1}}
\newlength{\mycolwidthD}
\newlength{\mycolwidthE}
\newlength{\mycolwidthC}
\settowidth{\mycolwidthD}{100.0}
\settowidth{\mycolwidthE}{100.0}
\settowidth{\mycolwidthC}{100.0}
\setlength{\mycolwidthD}{1.3\mycolwidthE}
\resizebox{0.9\linewidth}{!}{
\begin{tabular}{lc|*{3}{C{\mycolwidthD}}|*{3}{C{\mycolwidthE}}|*{4}{C{\mycolwidthC}}|c}
\toprule[1.5pt]
\multirow{2}{*}{\textbf{Method}} & \multirow{2}{*}{\textbf{Model}} 
& \multicolumn{3}{c|}{\textbf{Speech Content Extraction}} 
& \multicolumn{3}{c|}{\textbf{Audio Classification}} 
& \multicolumn{4}{c|}{\textbf{Speaker Recognition}} 
& \multirow{2}{*}{\textbf{Avg.}} \\
\cmidrule(lr){3-5} \cmidrule(lr){6-8} \cmidrule(lr){9-12}
& & \textbf{SER} & \textbf{SCR} & \textbf{ASR} 
& \textbf{SED} & \textbf{MC} & \textbf{ASC} 
& \textbf{SD} & \textbf{ER} & \textbf{SAR} & \textbf{SGR} & \\
\hline
\rowcolor{gray!20}
\multicolumn{13}{c}{\textit{{Vanilla}}} \\ 
& Phi-4-Multimodal & 18.4 & 69.3 & {92.7} & 55.1 & 46.7 & 23.4 & 26.4 & 27.3 & 26.6 & 91.1 & 47.7 \\
& Aero-1-Audio & 17.9 & 56.6 & {43.7} & 55.0 & 83.9 & 39.9 & 33.7 & 32.0 & 17.8 & 47.5 & 42.8\\ 
& Qwen2.5-Omni-3B & 25.2 & 82.3 & {94.7} & {70.2} & {100.0} & {69.3} & 67.3 & 39.6 & 29.1 & 97.2 & {\textbf{67.5}} \\
\rowcolor{mygray}
\multicolumn{13}{c}{{\textit{Light Token Pruning} \ $\fg{(\downarrow 30\%)}$}} \\ 
\multirow{3}{*}{Random} 
 & Phi-4-multimodal & 18.4 & 67.5 & 49.1 & 31.4 & 39.8 & 30.2 & 31.3 & 31.0 & 24.5 & 93.6 & 41.7 \\
 & Aero-1-Audio & 15.9 & 53.9 & 43.3 & 56.8 & 79.4 & 40.2 & 34.0 & 32.4 & 10.0 & 38.8 & 40.5 \\
 & Qwen2.5-Omni-3B & 26.5 & 90.3 & 88.4 & 71.1 & 97.5 & 69.7 & 72.0 & 38.4 & 28.6 & 95.7 & \textbf{67.8} \\
\cmidrule(lr){1-13}
\multirow{3}{*}{FastV} 
 & Phi-4-multimodal & 18.3 & 64.0 & 43.9 & 33.2 & 40.6 & 29.6 & 44.0 & 29.1 & 25.7 & 92.9 &  42.1\\
 & Aero-1-Audio & 19.7 & 57.0 & 37.5 & 57.0 & 78.8 & 41.0 & 42.1 & 32.2 & 9.2 & 39.2 & 41.4 \\
 & Qwen2.5-Omni-3B & 18.7 & 68.2 & 76.3 & 61.3 & 98.4 & 57.2 & 38.5 & 31.1 & 17.3 & 97.5 & \textbf{56.5}\\
\cmidrule(lr){1-13}
\multirow{3}{*}{DART} 
 & Phi-4-multimodal & 16.8 & 67.6 & 57.2 & 54.5 & 46.1 & 31.8 & 23.1 & 28.6 & 27.1 & 91.6 & 44.4 \\
 & Aero-1-Audio & 20.2 & 57.0 & 16.4 & 56.3 & 78.8 & 41.0 & 34.0 & 32.2 & 9.2 & 39.5 & 38.5 \\
 & Qwen2.5-Omni-3B & 23.2 & 74.2 & 81.4 & 73.1 & 97.6 & 72.5 & 42.2 & 37.1 & 23.0 & 48.7 & \textbf{57.3} \\
\cmidrule(lr){1-13}
\multirow{3}{*}{Frame (Ours)} 
 & Phi-4-multimodal & 17.7 & 64.4 & 63.4 & 31.4 & 32.6 & 29.0 & 30.6 & 31.0 & 27.4 & 92.4 & 42.0 \\
 & Aero-1-Audio & 15.6 & 53.7 & 43.4 & 54.3 & 82.5 & 39.8 & 34.4 & 32.1 & 8.1 & 37.3 & 40.1 \\
 & Qwen2.5-Omni-3B & 26.8 & 80.9 & 92.2 & 70.5 & 98.5 & 70.2 & 65.0 & 36.4 & 31.4 & 96.7 & \textbf{66.9}  \\
\hline
\rowcolor{mygray}
\multicolumn{13}{c}{{\textit{Medium Token Pruning} \ $\fg{(\downarrow 60\%)}$}} \\ 
\multirow{3}{*}{Random} 
 & Phi-4-multimodal & 18.7 & 61.6 & 7.9 & 30.3 & 27.4 & 30.6 & 36.8 & 29.4 & 20.5 & 91.2 & 35.4 \\
 & Aero-1-Audio & 12.1 & 49.7 & 34.9 & 54.6 & 78.2 & 41.3 & 42.5 & 34.5 & 8.8 & 34.0 & 39.1 \\
 & Qwen2.5-Omni-3B & 24.2 & 75.3 & 59.7 & 68.7 & 95.8 & 68.3 & 66.6 & 37.9 & 27.2 & 93.5 & \textbf{61.7} \\
\cmidrule(lr){1-13}
\multirow{3}{*}{FastV} 
 & Phi-4-multimodal & 26.1 & 52.8 & 0.0 & 32.5 & 28.2 & 30.3 & 25.6 & 28.0 & 22.2 & 89.4 & 33.5 \\
 & Aero-1-Audio & 20.3 & 54.2 & 30.4 & 58.0 & 80.2 & 44.5 & 34.2 & 33.4 & 9.1 & 34.5 & 39.9 \\
 & Qwen2.5-Omni-3B & 18.0 & 63.8 & 39.2 & 60.5 & 97.5 & 57.8 & 44.0 & 28.6 & 17.1 & 95.3 & \textbf{52.2}\\
\cmidrule(lr){1-13}
\multirow{3}{*}{DART} 
 & Phi-4-multimodal & 18.0 & 61.1 & 23.7 & 53.9 & 44.8 & 25.4 & 26.3 & 29.4 & 24.4 & 88.0 & 39.5 \\
 & Aero-1-Audio & 20.3 & 54.2 & 14.4 & 58.2 & 80.6 & 44.5 & 34.0 & 33.4 & 9.1 & 34.5 & 38.3 \\
 & Qwen2.5-Omni-3B & 23.1 & 64.6 & 62.8 & 71.9 & 99.1 & 73.4 & 38.3 & 37.6 & 28.1 & 46.0 & \textbf{54.5} \\
\cmidrule(lr){1-13}
\multirow{3}{*}{Frame (Ours)} 
 & Phi-4-multimodal & 23.8 & 59.0 & 23.3 & 31.1 & 28.5 & 30.0 & 20.6 & 30.1 & 22.3 & 87.8 & 35.6 \\
 & Aero-1-Audio & 14.0 & 51.7 & 42.5 & 56.4 & 80.9 & 41.1 & 35.0 & 34.5 & 9.1 & 33.3 & 39.9 \\
 & Qwen2.5-Omni-3B & 25.6 & 75.3 & 82.2 & 69.0 & 100.0 & 68.3 & 65.7 & 38.6 & 28.3 & 91.0 & \textbf{64.4} \\
\hline
\rowcolor{mygray}
\multicolumn{13}{c}{{\textit{Extreme Token Pruning} \ $\fg{(\downarrow 90\%)}$}} \\ 
\multirow{3}{*}{Random} 
 & Phi-4-multimodal & 18.7 & 35.3 & 0.0 & 29.3 & 20.6 & 29.6 & 41.9 & 33.4 & 11.1 & 67.5 & 28.7 \\
 & Aero-1-Audio & 10.1 & 47.6 & 5.1 & 43.4 & 70.3 & 44.9 & 47.5 & 32.2 & 14.6 & 33.3 & 34.9 \\
 & Qwen2.5-Omni-3B & 24.0 & 58.1 & 0.0 & 65.9 & 97.6 & 60.0 & 54.7 & 41.8 & 17.1 & 84.3 & \textbf{50.4} \\
\cmidrule(lr){1-13}
\multirow{3}{*}{FastV} 
 & Phi-4-multimodal & 23.4 & 43.0 & 0.0 & 27.6 & 30.1 & 29.3 & 46.3 & 24.9 & 17.3 & 82.9 & 32.5 \\
 & Aero-1-Audio & 18.0 & 50.6 & 8.3 & 55.8 & 69.0 & 45.0 & 38.2 & 26.3 & 16.8 & 33.6 & 36.2 \\
 & Qwen2.5-Omni-3B & 16.8 & 54.9 & 3.5 & 65.2 & 95.9 & 55.9 & 49.5 & 32.7 & 14.8 & 86.5 & \textbf{47.6}\\
\cmidrule(lr){1-13}
\multirow{3}{*}{DART} 
 & Phi-4-multimodal & 16.8 & 49.3 & 0.0 & 52.0 & 40.2 & 24.4 & 31.9 & 27.4 & 18.6 & 77.2 & 33.8 \\
 
 & Aero-1-Audio & 18.0 & 50.6 & 0.0 & 55.8 & 69.0 & 45.0 & 38.1 & 26.3 & 16.8 & 33.6 & 35.3 \\
 & Qwen2.5-Omni-3B & 17.3 & 54.1 & 62.9 & 66.8 & 99.1 & 69.3 & 25.6 & 42.6 & 22.1 & 52.8 & \textbf{51.3} \\
\cmidrule(lr){1-13}
\multirow{3}{*}{Frame (Ours)} 
 & Phi-4-multimodal & 24.6 & 36.8 & 0.0 & 28.4 & 25.0 & 28.1 & 34.8 & 30.1 & 12.1 & 66.6 & 28.7 \\
 & Aero-1-Audio & 9.7 & 48.9 & 3.1 & 43.8 & 73.0 & 44.0 & 54.2 & 33.2 & 16.1 & 33.3 & 35.9 \\
 & Qwen2.5-Omni-3B & 22.8 & 58.2 & 0.0 & 64.5 & 95.0 & 60.9 & 51.9 & 41.1 & 18.1 & 87.0 & \textbf{50.0} \\
\bottomrule[1.5pt]
\end{tabular}}
\label{tab:pruning_comparison_results}
\vspace{-0.9em}
\end{table}

\section{Efficiency Optimization for LALMs}
\label{method}
Processing extended audio sequences poses significant computational challenges for LALMs. A single 5-minute audio input can generate thousands of tokens, leading to quadratic memory growth and prohibitive inference latency (as shown in Table~\ref{tab:audio_models} of Appendix~\ref{app_sec:Encoding_Granularity}).
To address these bottlenecks, we systematically evaluate two complementary efficiency optimization strategies: \textbf{Token pruning}~\citep{liu2025shifting} during the prefilling stage and \textbf{KV-cache eviction}\footnote{\url{https://github.com/NVIDIA/kvpress}} during the decoding stage.




\vspace{-0.5em}
\subsection{Token Pruning and KV Cache Eviction}\label{token_pruning}
\vspace{-0.3em}
Processing long-form audio sequences poses substantial memory and latency challenges for LALMs. One-minute audio input is embedded into 1500 tokens, requiring massive KV-cache storage and significantly slow decoding, thus making deployment impractical without compression.
To address these bottlenecks, numerous approaches have emerged that directly reduce the number of tokens to improve inference efficiency.
We evaluate four token pruning methods and four KV cache eviction strategies on our long-audio benchmark.
Experiments are conducted on three open-source LALMs, including Qwen2.5-Omni-3B, Aero-1-Audio, and Phi-4-Multimodal.

\begin{figure*}[!h]
    \centering
    \includegraphics[width=1.0\linewidth]{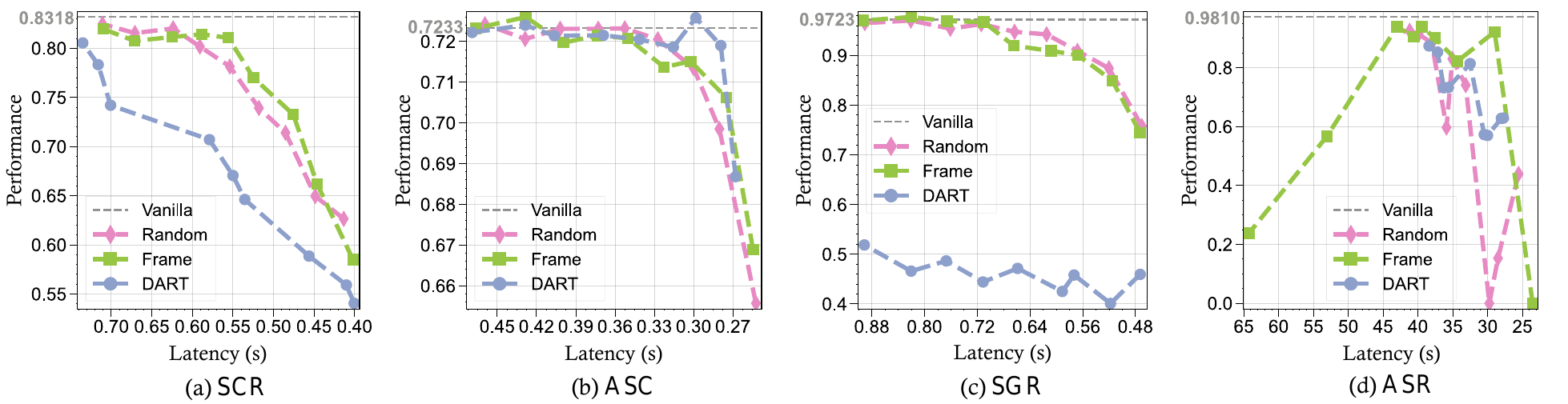}
    \vspace{-2em}
    \caption{Comparisons of latency and performance trade-off for the Qwen2.5-Omni-3B model under different token pruning strategies across four representative datasets. Frame consistently outperforms other methods across different latency constraints.}
    \label{fig:latency_performance}
    \vspace{-1em}
\end{figure*}

\begin{figure*}[!t]
    \centering
    \includegraphics[width=1.0\linewidth]{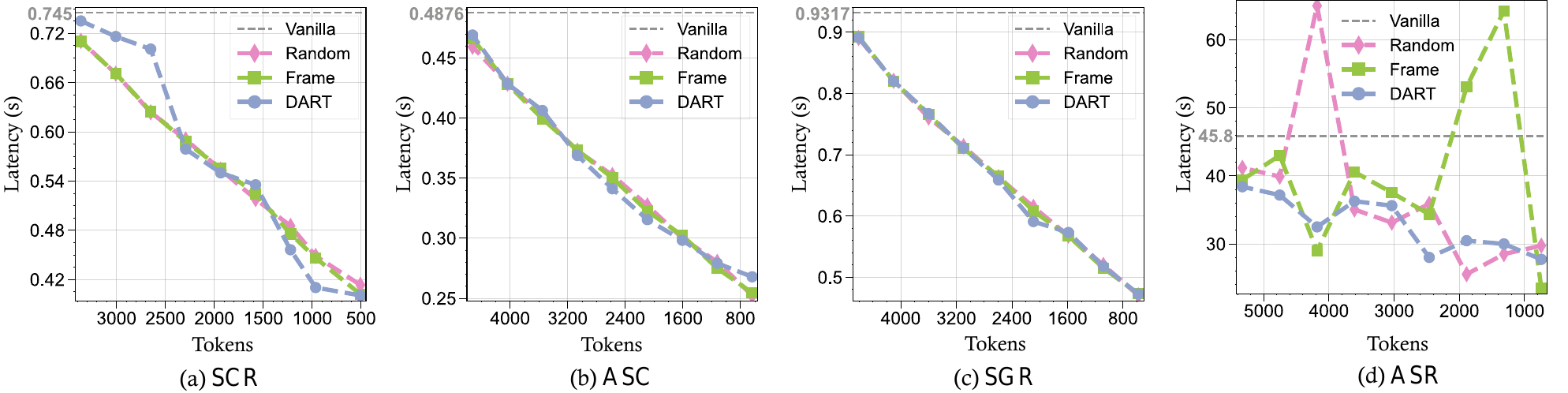}
    \vspace{-1.5em}
    \caption{Acceleration effects across token pruning strategies for the Qwen2.5-Omni-3B model under various token pruning strategies across four datasets.}
    \label{fig:tokens_latency}
    \vspace{-1.5em}
\end{figure*}
\noindent \textbf{Token pruning.} We compare four token pruning strategies on \benchname{}.
The baseline, Random pruning, discards tokens uniformly at random.
FastV~\citep{chen2024image} removes low-attention tokens, and DART~\citep{wen2025stop} applies redundancy-guided selection by discarding similar tokens.
However, due to the strongly sequential nature of acoustic signals, naive or purely attention-based pruning can inadvertently remove brief phonetic cues or transient events, leading to degraded recognition.
Unlike vision models, where redundancy often arises from spatial or semantic similarity, audio token redundancy primarily manifests as smooth temporal continuity.
Therefore, we additionally design \textbf{Frame} as a time-aligned token pruning strategy to preserve rare or short-lived acoustic events that other methods may discard, making it a scheme tailored to audio characteristics.

\noindent \textbf{KV-cache eviction.} We evaluate four eviction strategies under compression ratios of 30\%, 60\%, and 90\%. 
The baseline, Random eviction, uniformly removes cache entries, providing a lower bound on performance under cache pressure. 
KNorm~\citep{devoto2024simple} evicts tokens according to the L2 norm of their key vectors, based on the intuition that smaller norms contribute less to attention. 
TOVA~\citep{oren2024transformers} greedily discards tokens with minimal attention from the latest query by averaging attention weights across heads at each decoding step. 
Finally, SnapKV~\citep{li2024snapkv} retains high-attention tokens along with their neighbors using cumulative-attention scoring and 1D pooling-based clustering, preserving local semantic coherence while enabling efficient compression.

\begin{figure}[!h]
    \vspace{-3mm}
    \centering
    \includegraphics[width=1.0\linewidth]{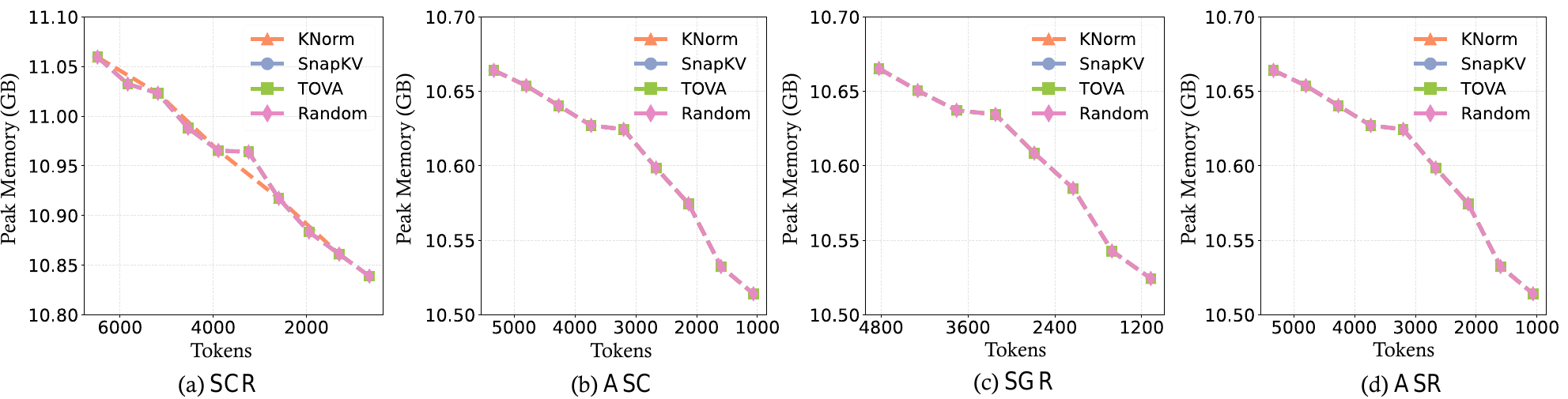}
    \vspace{-6mm}
    \caption{Peak GPU memory usage count for KV cache eviction policies applied to the Qwen2.5-Omni-3B model across four datasets, illustrating memory compression benefits during the prefilling stage for long-context audio inference.}
    \label{fig:cache_tokens}
    \vspace{-1em}
\end{figure}

\section{Results and Discussions}\label{sec:discussion}
\label{sec:results}
Our proposed \benchname{} provides a realism-oriented evaluation framework for LALMs, focusing on minute-scale recordings, diverse audio domains, and complex reasoning.
Results in Table~\ref{tab:AudioMarathon_results_categories_avg} reveal clear performance stratification among the 16 evaluated models.
\textbf{(i)} The best-performing model, Qwen2.5-Omni-7B, achieves an average F1-score of 70.5, whereas most open-source models cluster between 30 and 60, highlighting a substantial performance gap.
\textbf{(ii)} Closed-source models perform unevenly: all fail on long-audio emotion recognition and authenticity detection, with Gemini-2.5-Flash being the only one to exceed 30 in emotion recognition.
Even for authenticity detection, Qwen2.5-Omni reaches 72.3, while every closed-source model remains below 35.
\textbf{(iii)} Human evaluation yields an average F1-score of 87.6, clearly surpassing even the strongest models. The gap is most pronounced in Speaker Information Modeling tasks such as speech emotion recognition (SER) and speaker-based entity recognition (ER), where human performance (87.6) remains far above model scores (generally below 65). This pronounced weakness directly reflects the challenges of entity tracking and temporal reasoning discussed in the introduction, highlighting a concrete target for future improvement.
In summary, leading LALMs perform competitively on narrow, single-domain tasks but still struggle with long-form speech understanding.
These weaknesses point to priorities for future work: (i) richer pretraining on long-form and multi-source audio, (ii) improved multi-scale and spatially aware representations, and (iii) benchmarks that explicitly assess long-context extraction, audio scene classification, and efficiency under realistic conditions.

\noindent \textbf{What Challenges Do LALMs Face?}
Long-form Speech Entity Recognition (SER) and Speaker-Age Recognition are the most challenging tasks in our benchmark. 
The top model achieves an F1-score just above 30 on SER, with only Audio-Flamingo-3 exceeding 40, indicating a substantial capability gap among current LALMs.  
Memory consumption and inference latency pose additional challenges. 
The Figure~\ref{fig:cache_tokens} shows that cache eviction reduces peak memory during prefilling, though GPU usage remains high. 
Preserving the first output token maintains performance for most MCQs. 
Figure~\ref{fig:tokens_latency} indicates that processing all long-audio tokens at the second decoder layer is computationally expensive; reducing tokens to 10\% cuts processing time to 56\% of the baseline, achieving a 1.8$\times$ speedup.  
Unlike vision tokens, audio tokens encode strong temporal dependencies. Aggressive eviction can disrupt temporal coherence, especially in ASR, where every phoneme matters. 
Figure~\ref{fig:latency_performance} shows that improper pruning can produce repetitive tokens, increasing latency and lowering word accuracy. 
Overall, task-aware audio token compression provides significant runtime and memory savings and is essential for scaling long-audio LLM inference.
\noindent \textbf{Semantic vs. Acoustic: Mapping Capabilities in Long-Audio Tasks.}
\begin{figure}[!t]
    \centering
    \includegraphics[width=\linewidth]{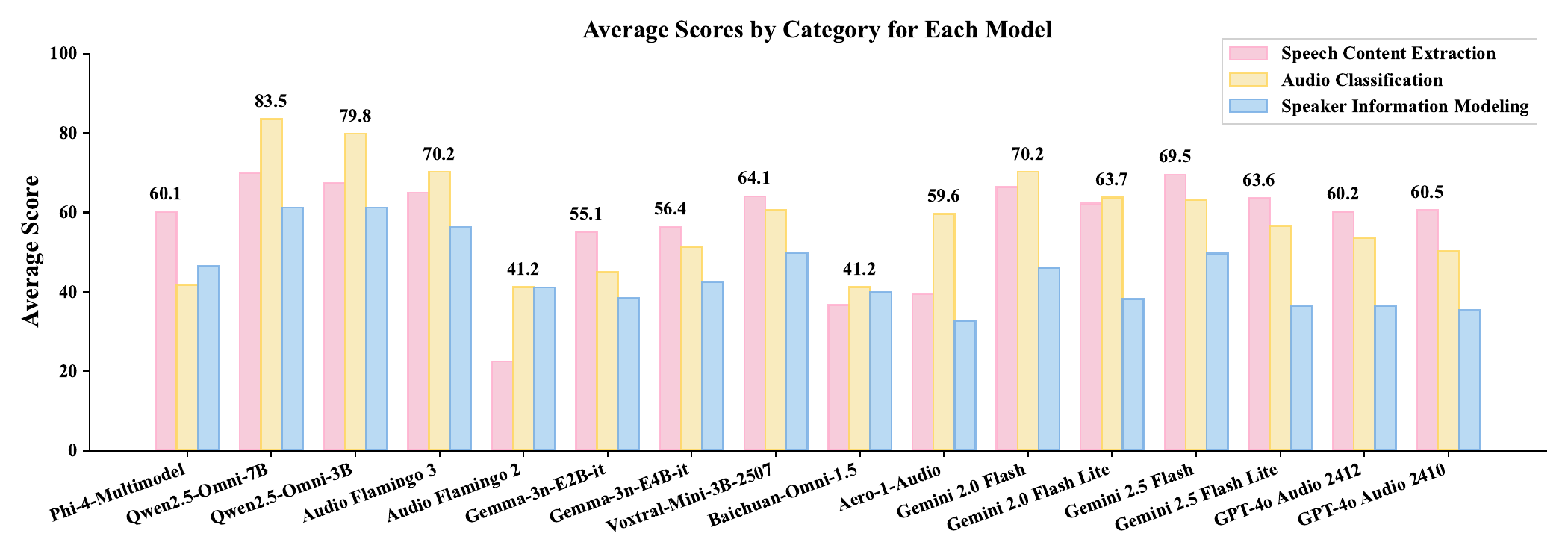}
    \vspace{-2em}
    \caption{Average F1-scores across the three main task categories: Speech Content Extraction, Audio Classification, and Speaker Information Modeling, underscoring the need for enhanced temporal reasoning in extended audio contexts.}
    \label{fig:category_scores}
    \vspace{-1.5em}
\end{figure}
Recent LALMs integrate acoustic and linguistic features within a single end-to-end model, enabling joint learning of cross-modal dependencies~\citep{peng2024survey}. 
In \benchname{}, ASR, Speech Content Reasoning (SCR), and SER are considered semantically sensitive tasks, while the remaining seven tasks are acoustically sensitive. 
Figure~\ref{fig:category_scores} shows that all closed-source models achieve around 60 F1-score on semantically sensitive tasks, reflecting their strength in extracting long-form audio content. 
The strongest model, Qwen2.5-Omni-7B, reaches 83.5, and the top four LALMs score above 70 on audio classification, indicating extensive training on classification tasks. 
In contrast, all models underperform on speaker-related tasks, failing to exceed a 65 F1-score, suggesting that speaker information modeling remains under-emphasized in current LALMs development.

\begin{figure}[h]
  \centering
  \vspace{-2mm}
  \includegraphics[width=\linewidth]{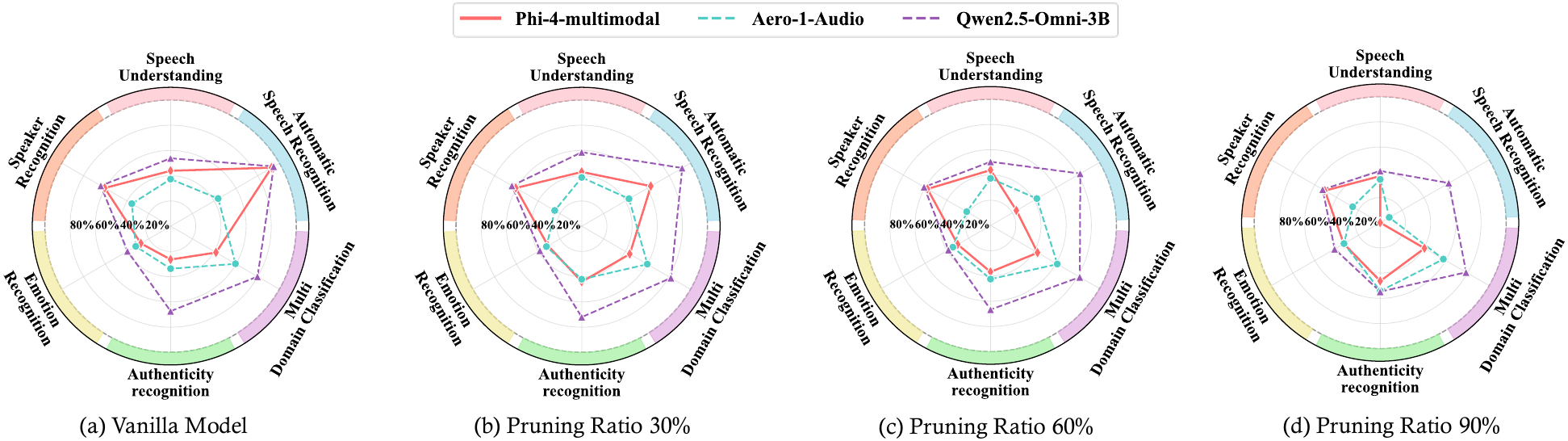}
  \caption{Performance comparison between Qwen2.5-Omni-3B, Phi-4-Mutimodal and Aero-1-Audio on six-degree capability under varying token pruning ratios.}
  \label{fig:radar_chart}
  \vspace{-0.8em}
\end{figure} 

\noindent \textbf{How do token pruning methods affect performance in LALMs?}
We evaluate six-degree capabilities: (i) \textit{Speech Understanding} (mean of SER and SCR), (ii) \textit{Speaker Recognition} (mean of SAR and SGR), (iii) \textit{Emotion Recognition} (ER), (iv) \textit{Authenticity Recognition} (SD), (v) \textit{Automatic Speech Recognition} (ASR), and (vi) \textit{Multi-Domain Classification} (mean of SED and MC), as shown in Figure ~\ref{fig:radar_chart}. For audio token pruning, the reported score is the maximum F1-score achieved across tested pruning settings.
\textbf{Qwen2.5-Omni-3B} benefits from token pruning, with Speech Understanding F1-score rising from 53.8 to 58.6, Multi-Domain Classification improving modestly (79.8 → 81.9), and Authenticity Recognition gaining about five points (67.3 → 72.0).
\textbf{Aero-1-Audio} is highly pruning-sensitive: its Speech Understanding F1-score loses roughly 5 points under light Frame pruning and over 15 under extreme settings, while DART exacerbates the degradation, resulting in total losses exceeding 25 points under the most aggressive configuration. Multi-Domain Classification remains comparatively robust, with only modest decreases under Frame. \textbf{Phi-4-Multimodal} shows a similar pattern: F1-score of Speech Understanding experiences moderate decline under light Frame pruning and severe degradation under extreme or DART-based pruning, whereas Multi-Domain Classification is better preserved with Frame than with DART.

Table~\ref{tab:pruning_comparison_results} highlights a task-dependent sensitivity gradient: temporally fine-grained tasks (ASR, Speech Understanding) degrade sharply when selective or attention-driven pruning removes temporally unique phonetic cues, while more global classification tasks (e.g., music detection) remain resilient even under aggressive compression. Mechanistically, DART’s redundancy-focused selection can discard rare temporal segments; Frame’s uniform windowed subsampling preserves coverage; FastV’s attention-based policy can misalign with linear temporal progression. Frame consistently attains the strongest speech-sensitive F1 scores across models and pruning ratios. These results argue for task-aware token pruning: redundancy-based strategies suit over-represented signals, Frame is a safe default for speech-centric workloads, and attention-driven pruning entails a tangible risk of temporal coherence disruption.

\section{Related Works}\label{sec:related_work}

\noindent\textbf{Large Audio Language Models.} 
The development of LALMs follows the broader shift toward multimodal language processing. Early audio models combined ASR and Text-to-Speech with text-based LLMs for audio-to-text tasks, but they suffered from error propagation and weak cross-modal fusion~\citep{ngiam2011multimodal,hinton2012deep,wang2017tacotron}. 
Self supervised speech representations, such as wav2vec 2.0~\citep{baevski2020wav2vec} and HuBERT~\citep{hsu2021hubert}, drove major progress and enabled models like Whisper~\citep{radford2023robust} and SpeechGPT~\citep{zhang2023speechgpt}. Recent instruction-tuned Audio LLMs, such as Phi-4-multimodal~\citep{abouelenin2025phi}, Freeze-Omni~\citep{wang2024freeze}, and Qwen2.5-Omni~\citep{xu2025qwen2}, unify audio and language within one framework and support tasks that span ASR, audio question answering, and audio understanding. As context windows grow~\citep{liu2025shifting}, these models are also moving toward longer audio inputs, with some reporting support for hours of audio.

\noindent\textbf{Audio LLM Benchmarks.} 
Benchmarking has evolved from task-specific datasets to broader frameworks that test multimodal and instruction-following abilities. Early datasets such as AudioSet~\citep{gemmeke2017audio}, LibriSpeech~\citep{panayotov2015librispeech}, ESC 50~\citep{piczak2015esc}, and FSD50K~\citep{fonseca2021fsd50k} focused on classification or ASR. SUPERB~\citep{yang2021superb} expanded speech evaluation with a broader task set. For audio language understanding, Clotho QA~\citep{drossos2020clotho} and AudioCaps~\citep{kim2019audiocaps} introduced question answering and captioning. More recent datasets, such as MMAU~\citep{sakshi2024mmau} and AIR Bench~\citep{yang2024air}, target instruction following and tri modal reasoning. Despite progress, few benchmarks directly test long audio comprehension or the efficiency of long sequence processing.

\noindent\textbf{Token Compression.}
Transformer-based models face memory and compute limits with long context and multimodal inputs. Two practical directions are KV cache eviction and token pruning~\citep{liu2025shifting, wen2025token, xiong2025prune2drive,yang2025efficientvla,chen2025variation,wang2025winning}. For KV cache eviction, SnapKV~\citep{li2024snapkv} clusters high attention tokens and stores centroids, H2O~\citep{zhang2023h2o} balances recent and salient tokens, and StreamingLLM~\citep{xiao2023efficient} uses fixed attention sinks and a sliding window for unbounded generation. In vision language models, token pruning reduces redundant visual tokens through architectural methods, such as Q-Former context tokens~\citep{li2023llama}, and through inference time methods, such as Token Merging~\citep{bolya2022token}, FastV~\citep{chen2024image}, SparseVLM~\citep{zhang2024sparsevlm}, and DART~\citep{wen2025stop}. While these methods are effective for vision or text tokens, research on audio token compression remains limited, and it is unknown how well these methods transfer to the audio modality.

\section{Conclusion}


We present \benchname{}, a comprehensive benchmark for LALMs that targets minute-scale speech, sound, and music inputs, spanning 10 representative audio tasks. Through these long-form scenarios, \benchname{} exposes fundamental challenges such as long-range dependency, temporal continuity, and source confusion. While existing LALMs perform well on short-range tasks like classification and reasoning, they struggle with long-span speech understanding and speaker analysis, revealing limitations in consistency and entity tracking, which exhibit a significant gap compared to human performance, providing a potential direction for future research. Moreover, the field of large audio models lacks attention to the efficiency of audio encoders, which, in practice, leads to substantial redundancy in audio tokens. Ultimately, \benchname{} provides a foundation for developing robust and efficient long-audio modeling.


\section*{Ethics Statement}
Our research introduces the \benchname{} to advance long-form audio understanding and inference efficiency in Large Audio Language Models (LALMs). We acknowledge the dual-use potential of this technology, which could be misused for generating deepfake audio or eroding privacy. We justify its public disclosure as a means to foster robustness and safety through transparent benchmarking and to highlight model limitations for proactive risk mitigation. We encourage future work to expand this effort with responsible practices across diverse languages and contexts. We hereby affirm that this work was conducted in strict compliance with academic ethics, with the primary goal of steering technological progress toward beneficial ends; any misuse of this research for unlawful or unethical purposes is unequivocally contrary to our principles.

\section*{Reproducibility Statement}
We are committed to ensuring the reproducibility of our results. To this end, we provide comprehensive details about our experimental setup and datasets. Specifically, all hyperparameters, model descriptions, pruning strategies, and evaluation protocols are specified in the main text. Additional analysis, including random baseline results, error analysis, encoding granularity of LALMs, and model details, is presented in the Appendix~\ref{sec:model_details}. We describe all datasets used in our benchmark (\benchname{}) in Appendix~\ref{sec:dataset_details}, including their construction and task definitions. Detailed motivation and implementation of pruning at the second layer are provided in Appendix~\ref{sec:additional_results}, where we explain its design as an early-stage compression mechanism for audio tokens. We report the performance of different pruning methods, random baselines, acceleration consistency, and model-specific results (e.g., Qwen2.5-Omni-3B) in the appendix for full transparency.  Appendix~\ref{sec:error_analysis} provides descriptions of all baseline models, including their architectures, training strategies, and modality support. Together, these efforts ensure that all experiments can be independently reproduced.
\bibliography{iclr2026_conference}

\begin{thebibliography}{82}
\providecommand{\natexlab}[1]{#1}
\providecommand{\url}[1]{\texttt{#1}}
\expandafter\ifx\csname urlstyle\endcsname\relax
  \providecommand{\doi}[1]{doi: #1}\else
  \providecommand{\doi}{doi: \begingroup \urlstyle{rm}\Url}\fi

\bibitem[Abdin et~al.(2024)Abdin, Aneja, Behl, Bubeck, Eldan, Gunasekar, Harrison, Hewett, Javaheripi, Kauffmann, et~al.]{phi4}
Marah Abdin, Jyoti Aneja, Harkirat Behl, S{\'e}bastien Bubeck, Ronen Eldan, Suriya Gunasekar, Michael Harrison, Russell~J Hewett, Mojan Javaheripi, Piero Kauffmann, et~al.
\newblock Phi-4 technical report.
\newblock \emph{arXiv preprint arXiv:2412.08905}, 2024.

\bibitem[Abouelenin et~al.(2025)Abouelenin, Ashfaq, Atkinson, Awadalla, Bach, Bao, Benhaim, Cai, Chaudhary, Chen, et~al.]{abouelenin2025phi}
Abdelrahman Abouelenin, Atabak Ashfaq, Adam Atkinson, Hany Awadalla, Nguyen Bach, Jianmin Bao, Alon Benhaim, Martin Cai, Vishrav Chaudhary, Congcong Chen, et~al.
\newblock Phi-4-mini technical report: Compact yet powerful multimodal language models via mixture-of-loras.
\newblock \emph{arXiv preprint arXiv:2503.01743}, 2025.

\bibitem[Ahia et~al.(2025)Ahia, Bartelds, Ahuja, Gonen, Hofmann, Arora, Li, Puttagunta, Adeyemi, Buchireddy, et~al.]{ahia2025blab}
Orevaoghene Ahia, Martijn Bartelds, Kabir Ahuja, Hila Gonen, Valentin Hofmann, Siddhant Arora, Shuyue~Stella Li, Vishal Puttagunta, Mofetoluwa Adeyemi, Charishma Buchireddy, et~al.
\newblock Blab: Brutally long audio bench.
\newblock \emph{arXiv preprint arXiv:2505.03054}, 2025.

\bibitem[Alayrac et~al.(2022)Alayrac, Donahue, Luc, Miech, Barr, Hasson, Lenc, Mensch, Millican, Reynolds, et~al.]{alayrac2022flamingo}
Jean-Baptiste Alayrac, Jeff Donahue, Pauline Luc, Antoine Miech, Iain Barr, Yana Hasson, Karel Lenc, Arthur Mensch, Katherine Millican, Malcolm Reynolds, et~al.
\newblock Flamingo: a visual language model for few-shot learning.
\newblock \emph{Advances in neural information processing systems}, 35:\penalty0 23716--23736, 2022.

\bibitem[Baevski et~al.(2020)Baevski, Zhou, Mohamed, and Auli]{baevski2020wav2vec}
Alexei Baevski, Yuhao Zhou, Abdelrahman Mohamed, and Michael Auli.
\newblock wav2vec 2.0: A framework for self-supervised learning of speech representations.
\newblock \emph{Advances in neural information processing systems}, 33:\penalty0 12449--12460, 2020.

\bibitem[Beltagy et~al.(2020)Beltagy, Peters, and Cohan]{beltagy2020longformer}
Iz~Beltagy, Matthew~E Peters, and Arman Cohan.
\newblock Longformer: The long-document transformer.
\newblock \emph{arXiv preprint arXiv:2004.05150}, 2020.

\bibitem[Bolya et~al.(2022)Bolya, Fu, Dai, Zhang, Feichtenhofer, and Hoffman]{bolya2022token}
Daniel Bolya, Cheng-Yang Fu, Xiaoliang Dai, Peizhao Zhang, Christoph Feichtenhofer, and Judy Hoffman.
\newblock Token merging: Your vit but faster.
\newblock \emph{arXiv preprint arXiv:2210.09461}, 2022.

\bibitem[Borsos et~al.(2023)Borsos, Marinier, Vincent, Kharitonov, Pietquin, Sharifi, Roblek, Teboul, Grangier, Tagliasacchi, et~al.]{borsos2023audiolm}
Zal{\'a}n Borsos, Rapha{\"e}l Marinier, Damien Vincent, Eugene Kharitonov, Olivier Pietquin, Matt Sharifi, Dominik Roblek, Olivier Teboul, David Grangier, Marco Tagliasacchi, et~al.
\newblock Audiolm: a language modeling approach to audio generation.
\newblock \emph{IEEE/ACM transactions on audio, speech, and language processing}, 31:\penalty0 2523--2533, 2023.

\bibitem[Chen et~al.(2025)Chen, Liu, Wen, Wang, Huang, and Chen]{chen2025variation}
Junjie Chen, Xuyang Liu, Zichen Wen, Yiyu Wang, Siteng Huang, and Honggang Chen.
\newblock Variation-aware vision token dropping for faster large vision-language models.
\newblock \emph{arXiv preprint arXiv:2509.01552}, 2025.

\bibitem[Chen et~al.(2024{\natexlab{a}})Chen, Zhao, Liu, Bai, Lin, Zhou, and Chang]{chen2024image}
Liang Chen, Haozhe Zhao, Tianyu Liu, Shuai Bai, Junyang Lin, Chang Zhou, and Baobao Chang.
\newblock An image is worth 1/2 tokens after layer 2: Plug-and-play inference acceleration for large vision-language models.
\newblock In \emph{European Conference on Computer Vision}, pp.\  19--35. Springer, 2024{\natexlab{a}}.

\bibitem[Chen et~al.(2024{\natexlab{b}})Chen, Du, Wen, Zhou, Cui, Weng, Tu, Wang, Tong, Huang, et~al.]{chen2024mj}
Zhaorun Chen, Yichao Du, Zichen Wen, Yiyang Zhou, Chenhang Cui, Zhenzhen Weng, Haoqin Tu, Chaoqi Wang, Zhengwei Tong, Qinglan Huang, et~al.
\newblock Mj-bench: Is your multimodal reward model really a good judge for text-to-image generation?
\newblock \emph{arXiv preprint arXiv:2407.04842}, 2024{\natexlab{b}}.

\bibitem[Chu et~al.(2023)Chu, Xu, Zhou, Yang, Zhang, Yan, Zhou, and Zhou]{chu2023qwen}
Yunfei Chu, Jin Xu, Xiaohuan Zhou, Qian Yang, Shiliang Zhang, Zhijie Yan, Chang Zhou, and Jingren Zhou.
\newblock Qwen-audio: Advancing universal audio understanding via unified large-scale audio-language models.
\newblock \emph{arXiv preprint arXiv:2311.07919}, 2023.

\bibitem[Chu et~al.(2024)Chu, Xu, Yang, Wei, Wei, Guo, Leng, Lv, He, Lin, et~al.]{chu2024qwen2}
Yunfei Chu, Jin Xu, Qian Yang, Haojie Wei, Xipin Wei, Zhifang Guo, Yichong Leng, Yuanjun Lv, Jinzheng He, Junyang Lin, et~al.
\newblock Qwen2-audio technical report.
\newblock \emph{arXiv preprint arXiv:2407.10759}, 2024.

\bibitem[Comanici et~al.(2025{\natexlab{a}})Comanici, Bieber, Schaekermann, Pasupat, Sachdeva, Dhillon, Blistein, Ram, Zhang, Rosen, et~al.]{comanici2025gemini}
Gheorghe Comanici, Eric Bieber, Mike Schaekermann, Ice Pasupat, Noveen Sachdeva, Inderjit Dhillon, Marcel Blistein, Ori Ram, Dan Zhang, Evan Rosen, et~al.
\newblock Gemini 2.5: Pushing the frontier with advanced reasoning, multimodality, long context, and next generation agentic capabilities.
\newblock \emph{arXiv preprint arXiv:2507.06261}, 2025{\natexlab{a}}.

\bibitem[Comanici et~al.(2025{\natexlab{b}})Comanici, Bieber, Schaekermann, Pasupat, Sachdeva, Dhillon, Blistein, Ram, Zhang, Rosen, et~al.]{gemini2.X}
Gheorghe Comanici, Eric Bieber, Mike Schaekermann, Ice Pasupat, Noveen Sachdeva, Inderjit Dhillon, Marcel Blistein, Ori Ram, Dan Zhang, Evan Rosen, et~al.
\newblock Gemini 2.5: Pushing the frontier with advanced reasoning, multimodality, long context, and next generation agentic capabilities.
\newblock \emph{arXiv preprint arXiv:2507.06261}, 2025{\natexlab{b}}.

\bibitem[Devoto et~al.(2024)Devoto, Zhao, Scardapane, and Minervini]{devoto2024simple}
Alessio Devoto, Yu~Zhao, Simone Scardapane, and Pasquale Minervini.
\newblock A simple and effective $ l\_2 $ norm-based strategy for kv cache compression.
\newblock \emph{arXiv preprint arXiv:2406.11430}, 2024.

\bibitem[Drossos et~al.(2020)Drossos, Lipping, and Virtanen]{drossos2020clotho}
Konstantinos Drossos, Samuel Lipping, and Tuomas Virtanen.
\newblock Clotho: An audio captioning dataset.
\newblock In \emph{ICASSP 2020-2020 IEEE International Conference on Acoustics, Speech and Signal Processing (ICASSP)}, pp.\  736--740. IEEE, 2020.

\bibitem[Fonseca et~al.(2021)Fonseca, Favory, Pons, Font, and Serra]{fonseca2021fsd50k}
Eduardo Fonseca, Xavier Favory, Jordi Pons, Frederic Font, and Xavier Serra.
\newblock Fsd50k: an open dataset of human-labeled sound events.
\newblock \emph{IEEE/ACM Transactions on Audio, Speech, and Language Processing}, 30:\penalty0 829--852, 2021.

\bibitem[Gemmeke et~al.(2017)Gemmeke, Ellis, Freedman, Jansen, Lawrence, Moore, Plakal, and Ritter]{gemmeke2017audio}
Jort~F Gemmeke, Daniel~PW Ellis, Dylan Freedman, Aren Jansen, Wade Lawrence, R~Channing Moore, Manoj Plakal, and Marvin Ritter.
\newblock Audio set: An ontology and human-labeled dataset for audio events.
\newblock In \emph{2017 IEEE international conference on acoustics, speech and signal processing (ICASSP)}, pp.\  776--780. IEEE, 2017.

\bibitem[Ghosh et~al.(2025)Ghosh, Kong, Kumar, Sakshi, Kim, Ping, Valle, Manocha, and Catanzaro]{ghosh2025audio}
Sreyan Ghosh, Zhifeng Kong, Sonal Kumar, S~Sakshi, Jaehyeon Kim, Wei Ping, Rafael Valle, Dinesh Manocha, and Bryan Catanzaro.
\newblock Audio flamingo 2: An audio-language model with long-audio understanding and expert reasoning abilities.
\newblock \emph{arXiv preprint arXiv:2503.03983}, 2025.

\bibitem[Goel et~al.(2025)Goel, Ghosh, Kim, Kumar, Kong, Lee, Yang, Duraiswami, Manocha, Valle, et~al.]{goel2025audio}
Arushi Goel, Sreyan Ghosh, Jaehyeon Kim, Sonal Kumar, Zhifeng Kong, Sang-gil Lee, Chao-Han~Huck Yang, Ramani Duraiswami, Dinesh Manocha, Rafael Valle, et~al.
\newblock Audio flamingo 3: Advancing audio intelligence with fully open large audio language models.
\newblock \emph{arXiv preprint arXiv:2507.08128}, 2025.

\bibitem[Hechmi et~al.(2021)Hechmi, Trong, Hautam{\"a}ki, and Kinnunen]{hechmi2021voxceleb}
Khaled Hechmi, Trung~Ngo Trong, Ville Hautam{\"a}ki, and Tomi Kinnunen.
\newblock Voxceleb enrichment for age and gender recognition.
\newblock In \emph{2021 IEEE Automatic Speech Recognition and Understanding Workshop (ASRU)}, pp.\  687--693. IEEE, 2021.

\bibitem[Heittola et~al.(2019)Heittola, Mesaros, and Virtanen]{heittola2019tau}
Toni Heittola, Annamaria Mesaros, and Tuomas Virtanen.
\newblock {TAU} urban acoustic scenes 2019, development dataset.
\newblock Zenodo, March 2019.
\newblock URL \url{https://doi.org/10.5281/zenodo.2589280}.
\newblock Version 1.0.

\bibitem[Hinton et~al.(2012)Hinton, Deng, Yu, Dahl, Mohamed, Jaitly, Senior, Vanhoucke, Nguyen, Sainath, et~al.]{hinton2012deep}
Geoffrey Hinton, Li~Deng, Dong Yu, George~E Dahl, Abdel-rahman Mohamed, Navdeep Jaitly, Andrew Senior, Vincent Vanhoucke, Patrick Nguyen, Tara~N Sainath, et~al.
\newblock Deep neural networks for acoustic modeling in speech recognition: The shared views of four research groups.
\newblock \emph{IEEE Signal processing magazine}, 29\penalty0 (6):\penalty0 82--97, 2012.

\bibitem[Hsu et~al.(2021)Hsu, Bolte, Tsai, Lakhotia, Salakhutdinov, and Mohamed]{hsu2021hubert}
Wei-Ning Hsu, Benjamin Bolte, Yao-Hung~Hubert Tsai, Kushal Lakhotia, Ruslan Salakhutdinov, and Abdelrahman Mohamed.
\newblock Hubert: Self-supervised speech representation learning by masked prediction of hidden units.
\newblock \emph{IEEE/ACM transactions on audio, speech, and language processing}, 29:\penalty0 3451--3460, 2021.

\bibitem[Hurst et~al.(2024)Hurst, Lerer, Goucher, Perelman, Ramesh, Clark, Ostrow, Welihinda, Hayes, Radford, et~al.]{hurst2024gpt}
Aaron Hurst, Adam Lerer, Adam~P Goucher, Adam Perelman, Aditya Ramesh, Aidan Clark, AJ~Ostrow, Akila Welihinda, Alan Hayes, Alec Radford, et~al.
\newblock Gpt-4o system card.
\newblock \emph{arXiv preprint arXiv:2410.21276}, 2024.

\bibitem[Jin et~al.(2025)Jin, Wang, Gao, Wen, Qi, Liu, and Zhang]{jin2025thinking}
Xiangqi Jin, Yuxuan Wang, Yifeng Gao, Zichen Wen, Biqing Qi, Dongrui Liu, and Linfeng Zhang.
\newblock Thinking inside the mask: In-place prompting in diffusion llms.
\newblock \emph{arXiv preprint arXiv:2508.10736}, 2025.

\bibitem[Kang et~al.(2025)Kang, Wen, Wen, Ye, Li, Feng, Zhou, Wang, Lin, Zhang, et~al.]{kang2025legion}
Hengrui Kang, Siwei Wen, Zichen Wen, Junyan Ye, Weijia Li, Peilin Feng, Baichuan Zhou, Bin Wang, Dahua Lin, Linfeng Zhang, et~al.
\newblock Legion: Learning to ground and explain for synthetic image detection.
\newblock \emph{arXiv preprint arXiv:2503.15264}, 2025.

\bibitem[Kim et~al.(2019)Kim, Kim, Lee, and Kim]{kim2019audiocaps}
Chris~Dongjoo Kim, Byeongchang Kim, Hyunmin Lee, and Gunhee Kim.
\newblock Audiocaps: Generating captions for audios in the wild.
\newblock In \emph{Proceedings of the 2019 Conference of the North American Chapter of the Association for Computational Linguistics: Human Language Technologies, Volume 1 (Long and Short Papers)}, pp.\  119--132, 2019.

\bibitem[Lai et~al.(2017)Lai, Xie, Liu, Yang, and Hovy]{lai2017race}
Guokun Lai, Qizhe Xie, Hanxiao Liu, Yiming Yang, and Eduard Hovy.
\newblock Race: Large-scale reading comprehension dataset from examinations.
\newblock \emph{arXiv preprint arXiv:1704.04683}, 2017.

\bibitem[Li et~al.(2025{\natexlab{a}})Li, Loy, Fanyi, Jingkang, Kaichen, Kairui, Minh, Quang, Ba, Shuai, Yezhen, and Ziwei]{li2025aero}
Bo~Li, Chen~Change Loy, Pu~Fanyi, Yang Jingkang, Zhang Kaichen, Hu~Kairui, Thang~Luu Minh, Trung~Nguyen Quang, Cong~Pham Ba, Liu Shuai, Wang Yezhen, and Liu Ziwei.
\newblock Aero: Audio-enhanced large language models.
\newblock 2025{\natexlab{a}}.
\newblock URL \url{https://www.lmms-lab.com/posts/aero_audio/}.

\bibitem[Li et~al.(2023)Li, Li, Savarese, and Hoi]{li2023blip}
Junnan Li, Dongxu Li, Silvio Savarese, and Steven Hoi.
\newblock Blip-2: Bootstrapping language-image pre-training with frozen image encoders and large language models.
\newblock In \emph{International conference on machine learning}, pp.\  19730--19742. PMLR, 2023.

\bibitem[Li et~al.(2025{\natexlab{b}})Li, Chen, Li, Liu, Wen, Shan, Liu, Liu, Liu, Wang, et~al.]{li2025tactic}
Weiya Li, Junjie Chen, Bei Li, Boyang Liu, Zichen Wen, Nuanqiao Shan, Xiaoqian Liu, Anping Liu, Huajie Liu, Youyan Wang, et~al.
\newblock Tactic: Translation agents with cognitive-theoretic interactive collaboration.
\newblock \emph{arXiv preprint arXiv:2506.08403}, 2025{\natexlab{b}}.

\bibitem[Li et~al.(2025{\natexlab{c}})Li, Liu, Zhang, Chen, Li, Li, Liu, Ming, Dong, Pan, et~al.]{li2025baichuan}
Yadong Li, Jun Liu, Tao Zhang, Song Chen, Tianpeng Li, Zehuan Li, Lijun Liu, Lingfeng Ming, Guosheng Dong, Da~Pan, et~al.
\newblock Baichuan-omni-1.5 technical report.
\newblock \emph{arXiv preprint arXiv:2501.15368}, 2025{\natexlab{c}}.

\bibitem[Li et~al.(2024{\natexlab{a}})Li, Wang, and Jia]{li2023llama}
Yanwei Li, Chengyao Wang, and Jiaya Jia.
\newblock {LLaMA-VID}: An image is worth 2 tokens in large language models.
\newblock In \emph{Proceedings of the IEEE/CVF Conference on Computer Vision and Pattern Recognition}, 2024{\natexlab{a}}.

\bibitem[Li et~al.(2024{\natexlab{b}})Li, Huang, Yang, Venkitesh, Locatelli, Ye, Cai, Lewis, and Chen]{li2024snapkv}
Yuhong Li, Yingbing Huang, Bowen Yang, Bharat Venkitesh, Acyr Locatelli, Hanchen Ye, Tianle Cai, Patrick Lewis, and Deming Chen.
\newblock Snapkv: Llm knows what you are looking for before generation.
\newblock \emph{Advances in Neural Information Processing Systems}, 37:\penalty0 22947--22970, 2024{\natexlab{b}}.

\bibitem[Liu et~al.(2023)Liu, Li, Wu, and Lee]{liu2023visual}
Haotian Liu, Chunyuan Li, Qingyang Wu, and Yong~Jae Lee.
\newblock Visual instruction tuning.
\newblock \emph{Advances in neural information processing systems}, 36:\penalty0 34892--34916, 2023.

\bibitem[Liu et~al.(2025)Liu, Wen, Wang, Chen, Tao, Wang, Jin, Zou, Wang, Liao, et~al.]{liu2025shifting}
Xuyang Liu, Zichen Wen, Shaobo Wang, Junjie Chen, Zhishan Tao, Yubo Wang, Xiangqi Jin, Chang Zou, Yiyu Wang, Chenfei Liao, et~al.
\newblock Shifting ai efficiency from model-centric to data-centric compression.
\newblock \emph{arXiv preprint arXiv:2505.19147}, 2025.

\bibitem[Ma et~al.(2025)Ma, Ma, Zhu, Yang, Chao, Xu, Chen, Chen, Chen, Cong, et~al.]{ma2025mmar}
Ziyang Ma, Yinghao Ma, Yanqiao Zhu, Chen Yang, Yi-Wen Chao, Ruiyang Xu, Wenxi Chen, Yuanzhe Chen, Zhuo Chen, Jian Cong, et~al.
\newblock Mmar: A challenging benchmark for deep reasoning in speech, audio, music, and their mix.
\newblock \emph{arXiv preprint arXiv:2505.13032}, 2025.

\bibitem[Nagrani et~al.(2017)Nagrani, Chung, and Zisserman]{nagrani2017voxceleb}
Arsha Nagrani, Joon~Son Chung, and Andrew Zisserman.
\newblock Voxceleb: a large-scale speaker identification dataset.
\newblock \emph{arXiv preprint arXiv:1706.08612}, 2017.

\bibitem[Nayak(2025)]{nayak2025kokoro}
Aryan Nayak.
\newblock Kokoro: An accessible text-to-speech application for visually impaired students.
\newblock \emph{No, this is the first time. I'm ever publishing a research paper}, 2025.

\bibitem[Ngiam et~al.(2011)Ngiam, Khosla, Kim, Nam, Lee, Ng, et~al.]{ngiam2011multimodal}
Jiquan Ngiam, Aditya Khosla, Mingyu Kim, Juhan Nam, Honglak Lee, Andrew~Y Ng, et~al.
\newblock Multimodal deep learning.
\newblock In \emph{ICML}, volume~11, pp.\  689--696, 2011.

\bibitem[Nie et~al.(2025)Nie, Zhu, You, Zhang, Ou, Hu, Zhou, Lin, Wen, and Li]{nie2025large}
Shen Nie, Fengqi Zhu, Zebin You, Xiaolu Zhang, Jingyang Ou, Jun Hu, Jun Zhou, Yankai Lin, Ji-Rong Wen, and Chongxuan Li.
\newblock Large language diffusion models.
\newblock \emph{arXiv preprint arXiv:2502.09992}, 2025.

\bibitem[Oren et~al.(2024)Oren, Hassid, Yarden, Adi, and Schwartz]{oren2024transformers}
Matanel Oren, Michael Hassid, Nir Yarden, Yossi Adi, and Roy Schwartz.
\newblock Transformers are multi-state rnns.
\newblock \emph{arXiv preprint arXiv:2401.06104}, 2024.

\bibitem[Panayotov et~al.(2015)Panayotov, Chen, Povey, and Khudanpur]{panayotov2015librispeech}
Vassil Panayotov, Guoguo Chen, Daniel Povey, and Sanjeev Khudanpur.
\newblock Librispeech: an asr corpus based on public domain audio books.
\newblock In \emph{2015 IEEE international conference on acoustics, speech and signal processing (ICASSP)}, pp.\  5206--5210. IEEE, 2015.

\bibitem[Park et~al.(2024)Park, Salazar, Jansen, Kinoshita, Ro, and Skerry{-}Ryan]{park2024long}
Se~Jin Park, Julian Salazar, Aren Jansen, Keisuke Kinoshita, Yong~Man Ro, and R.~J. Skerry{-}Ryan.
\newblock Long-form speech generation with spoken language models.
\newblock \emph{CoRR}, abs/2412.18603, 2024.

\bibitem[Peng et~al.(2024)Peng, Wang, Xi, Li, Zhang, and Yu]{peng2024survey}
Jing Peng, Yucheng Wang, Yu~Xi, Xu~Li, Xizhuo Zhang, and Kai Yu.
\newblock A survey on speech large language models.
\newblock \emph{arXiv e-prints}, pp.\  arXiv--2410, 2024.

\bibitem[Piczak(2015)]{piczak2015esc}
Karol~J Piczak.
\newblock Esc: Dataset for environmental sound classification.
\newblock In \emph{Proceedings of the 23rd ACM international conference on Multimedia}, pp.\  1015--1018, 2015.

\bibitem[Radford et~al.(2023)Radford, Kim, Xu, Brockman, McLeavey, and Sutskever]{radford2023robust}
Alec Radford, Jong~Wook Kim, Tao Xu, Greg Brockman, Christine McLeavey, and Ilya Sutskever.
\newblock Robust speech recognition via large-scale weak supervision.
\newblock In \emph{International conference on machine learning}, pp.\  28492--28518. PMLR, 2023.

\bibitem[Sager et~al.(2019)Sager, Shankar, Reinhold, and Venkataraman]{sager2019vesus}
Jacob Sager, Ravi Shankar, Jacob Reinhold, and Archana Venkataraman.
\newblock Vesus: A crowd-annotated database to study emotion production and perception in spoken english.
\newblock In \emph{Interspeech}, pp.\  316--320, 2019.

\bibitem[Sakshi et~al.(2024)Sakshi, Tyagi, Kumar, Seth, Selvakumar, Nieto, Duraiswami, Ghosh, and Manocha]{sakshi2024mmau}
S~Sakshi, Utkarsh Tyagi, Sonal Kumar, Ashish Seth, Ramaneswaran Selvakumar, Oriol Nieto, Ramani Duraiswami, Sreyan Ghosh, and Dinesh Manocha.
\newblock Mmau: A massive multi-task audio understanding and reasoning benchmark.
\newblock \emph{arXiv preprint arXiv:2410.19168}, 2024.

\bibitem[Shon et~al.(2022)Shon, Pasad, Wu, Brusco, Artzi, Livescu, and Han]{shon2022slue}
Suwon Shon, Ankita Pasad, Felix Wu, Pablo Brusco, Yoav Artzi, Karen Livescu, and Kyu~J Han.
\newblock Slue: New benchmark tasks for spoken language understanding evaluation on natural speech.
\newblock In \emph{ICASSP 2022-2022 IEEE International Conference on Acoustics, Speech and Signal Processing (ICASSP)}, pp.\  7927--7931. IEEE, 2022.

\bibitem[Team(2025)]{gemma_3n_2025}
Gemma Team.
\newblock Gemma 3n.
\newblock 2025.
\newblock URL \url{https://ai.google.dev/gemma/docs/gemma-3n}.

\bibitem[Turpault et~al.(2019)Turpault, Serizel, Shah, and Salamon]{turpault2019sound}
Nicolas Turpault, Romain Serizel, Ankit~Parag Shah, and Justin Salamon.
\newblock Sound event detection in domestic environments with weakly labeled data and soundscape synthesis.
\newblock In \emph{Workshop on Detection and Classification of Acoustic Scenes and Events}, 2019.

\bibitem[Tzanetakis \& Cook(2002)Tzanetakis and Cook]{tzanetakis2002musical}
George Tzanetakis and Perry Cook.
\newblock Musical genre classification of audio signals.
\newblock \emph{IEEE Transactions on speech and audio processing}, 10\penalty0 (5):\penalty0 293--302, 2002.

\bibitem[Vaswani et~al.(2017)Vaswani, Shazeer, Parmar, Uszkoreit, Jones, Gomez, Kaiser, and Polosukhin]{vaswani2017attention}
Ashish Vaswani, Noam Shazeer, Niki Parmar, Jakob Uszkoreit, Llion Jones, Aidan~N Gomez, {\L}ukasz Kaiser, and Illia Polosukhin.
\newblock Attention is all you need.
\newblock \emph{Advances in neural information processing systems}, 30, 2017.

\bibitem[Wang et~al.(2024{\natexlab{a}})Wang, Zou, Lin, Sun, Liu, Zhang, Liu, Aw, and Chen]{wang2024audiobench}
Bin Wang, Xunlong Zou, Geyu Lin, Shuo Sun, Zhuohan Liu, Wenyu Zhang, Zhengyuan Liu, AiTi Aw, and Nancy~F Chen.
\newblock Audiobench: A universal benchmark for audio large language models.
\newblock \emph{arXiv preprint arXiv:2406.16020}, 2024{\natexlab{a}}.

\bibitem[Wang et~al.(2025{\natexlab{a}})Wang, Wu, Li, Yang, Chen, Zhang, and Meng]{wang2025mmsu}
Dingdong Wang, Jincenzi Wu, Junan Li, Dongchao Yang, Xueyuan Chen, Tianhua Zhang, and Helen Meng.
\newblock Mmsu: A massive multi-task spoken language understanding and reasoning benchmark.
\newblock \emph{arXiv preprint arXiv:2506.04779}, 2025{\natexlab{a}}.

\bibitem[Wang et~al.(2025{\natexlab{b}})Wang, Wang, Zhang, Wang, Min, Wen, Huang, Jiang, Lin, Liu, et~al.]{wang2025winning}
Shaobo Wang, Jiaming Wang, Jiajun Zhang, Cong Wang, Yue Min, Zichen Wen, Fei Huang, Huiqiang Jiang, Junyang Lin, Dayiheng Liu, et~al.
\newblock Winning the pruning gamble: A unified approach to joint sample and token pruning for efficient supervised fine-tuning.
\newblock \emph{arXiv preprint arXiv:2509.23873}, 2025{\natexlab{b}}.

\bibitem[Wang et~al.(2024{\natexlab{b}})Wang, Li, Fu, Shen, Xie, Li, Sun, and Ma]{wang2024freeze}
Xiong Wang, Yangze Li, Chaoyou Fu, Yunhang Shen, Lei Xie, Ke~Li, Xing Sun, and Long Ma.
\newblock Freeze-omni: A smart and low latency speech-to-speech dialogue model with frozen llm.
\newblock \emph{arXiv preprint arXiv:2411.00774}, 2024{\natexlab{b}}.

\bibitem[Wang et~al.(2017)Wang, Skerry-Ryan, Stanton, Wu, Weiss, Jaitly, Yang, Xiao, Chen, Bengio, et~al.]{wang2017tacotron}
Yuxuan Wang, RJ~Skerry-Ryan, Daisy Stanton, Yonghui Wu, Ron~J Weiss, Navdeep Jaitly, Zongheng Yang, Ying Xiao, Zhifeng Chen, Samy Bengio, et~al.
\newblock Tacotron: A fully end-to-end text-to-speech synthesis model.
\newblock \emph{arXiv preprint arXiv:1703.10135}, 164, 2017.

\bibitem[Weck et~al.(2024)Weck, Manco, Benetos, Quinton, Fazekas, and Bogdanov]{weck2024muchomusic}
Benno Weck, Ilaria Manco, Emmanouil Benetos, Elio Quinton, George Fazekas, and Dmitry Bogdanov.
\newblock Muchomusic: Evaluating music understanding in multimodal audio-language models.
\newblock \emph{arXiv preprint arXiv:2408.01337}, 2024.

\bibitem[Wen et~al.(2024)Wen, Guo, and Zhang]{wen2024aidbench}
Zichen Wen, Dadi Guo, and Huishuai Zhang.
\newblock Aidbench: A benchmark for evaluating the authorship identification capability of large language models.
\newblock \emph{arXiv preprint arXiv:2411.13226}, 2024.

\bibitem[Wen et~al.(2025{\natexlab{a}})Wen, Gao, Li, He, and Zhang]{wen2025token}
Zichen Wen, Yifeng Gao, Weijia Li, Conghui He, and Linfeng Zhang.
\newblock Token pruning in multimodal large language models: Are we solving the right problem?
\newblock \emph{arXiv preprint arXiv:2502.11501}, 2025{\natexlab{a}}.

\bibitem[Wen et~al.(2025{\natexlab{b}})Wen, Gao, Wang, Zhang, Zhang, Li, He, and Zhang]{wen2025stop}
Zichen Wen, Yifeng Gao, Shaobo Wang, Junyuan Zhang, Qintong Zhang, Weijia Li, Conghui He, and Linfeng Zhang.
\newblock Stop looking for important tokens in multimodal language models: Duplication matters more.
\newblock \emph{arXiv preprint arXiv:2502.11494}, 2025{\natexlab{b}}.

\bibitem[Wen et~al.(2025{\natexlab{c}})Wen, Qu, Liu, Liu, Wu, Yang, Jin, Xu, Liu, Li, et~al.]{wen2025devil}
Zichen Wen, Jiashu Qu, Dongrui Liu, Zhiyuan Liu, Ruixi Wu, Yicun Yang, Xiangqi Jin, Haoyun Xu, Xuyang Liu, Weijia Li, et~al.
\newblock The devil behind the mask: An emergent safety vulnerability of diffusion llms.
\newblock \emph{arXiv preprint arXiv:2507.11097}, 2025{\natexlab{c}}.

\bibitem[Wen et~al.(2025{\natexlab{d}})Wen, Wang, Zhou, Zhang, Zhang, Gao, Chen, Wang, Li, He, et~al.]{wen2025efficient}
Zichen Wen, Shaobo Wang, Yufa Zhou, Junyuan Zhang, Qintong Zhang, Yifeng Gao, Zhaorun Chen, Bin Wang, Weijia Li, Conghui He, et~al.
\newblock Efficient multi-modal large language models via progressive consistency distillation.
\newblock \emph{arXiv preprint arXiv:2510.00515}, 2025{\natexlab{d}}.

\bibitem[Xiao et~al.(2023)Xiao, Tian, Chen, Han, and Lewis]{xiao2023efficient}
Guangxuan Xiao, Yuandong Tian, Beidi Chen, Song Han, and Mike Lewis.
\newblock Efficient streaming language models with attention sinks.
\newblock \emph{arXiv preprint arXiv:2309.17453}, 2023.

\bibitem[Xiong et~al.(2025)Xiong, Wen, Gu, Liu, Zhang, Kang, Yang, Zhang, Li, He, et~al.]{xiong2025prune2drive}
Minhao Xiong, Zichen Wen, Zhuangcheng Gu, Xuyang Liu, Rui Zhang, Hengrui Kang, Jiabing Yang, Junyuan Zhang, Weijia Li, Conghui He, et~al.
\newblock Prune2drive: A plug-and-play framework for accelerating vision-language models in autonomous driving.
\newblock \emph{arXiv preprint arXiv:2508.13305}, 2025.

\bibitem[Xu et~al.(2025{\natexlab{a}})Xu, Guo, He, Hu, He, Bai, Chen, Wang, Fan, Dang, et~al.]{qwen2.5-omni}
Jin Xu, Zhifang Guo, Jinzheng He, Hangrui Hu, Ting He, Shuai Bai, Keqin Chen, Jialin Wang, Yang Fan, Kai Dang, et~al.
\newblock Qwen2. 5-omni technical report.
\newblock \emph{arXiv preprint arXiv:2503.20215}, 2025{\natexlab{a}}.

\bibitem[Xu et~al.(2025{\natexlab{b}})Xu, Guo, He, Hu, He, Bai, Chen, Wang, Fan, Dang, et~al.]{xu2025qwen2}
Jin Xu, Zhifang Guo, Jinzheng He, Hangrui Hu, Ting He, Shuai Bai, Keqin Chen, Jialin Wang, Yang Fan, Kai Dang, et~al.
\newblock Qwen2. 5-omni technical report.
\newblock \emph{arXiv preprint arXiv:2503.20215}, 2025{\natexlab{b}}.

\bibitem[Yang et~al.(2024)Yang, Xu, Liu, Chu, Jiang, Zhou, Leng, Lv, Zhao, Zhou, et~al.]{yang2024air}
Qian Yang, Jin Xu, Wenrui Liu, Yunfei Chu, Ziyue Jiang, Xiaohuan Zhou, Yichong Leng, Yuanjun Lv, Zhou Zhao, Chang Zhou, et~al.
\newblock Air-bench: Benchmarking large audio-language models via generative comprehension.
\newblock \emph{arXiv preprint arXiv:2402.07729}, 2024.

\bibitem[Yang et~al.(2021)Yang, Chi, Chuang, Lai, Lakhotia, Lin, Liu, Shi, Chang, Lin, et~al.]{yang2021superb}
Shu-wen Yang, Po-Han Chi, Yung-Sung Chuang, Cheng-I~Jeff Lai, Kushal Lakhotia, Yist~Y Lin, Andy~T Liu, Jiatong Shi, Xuankai Chang, Guan-Ting Lin, et~al.
\newblock Superb: Speech processing universal performance benchmark.
\newblock \emph{arXiv preprint arXiv:2105.01051}, 2021.

\bibitem[Yang et~al.(2025)Yang, Wang, Wen, Zhongwei, Zou, Zhang, Wen, and Zhang]{yang2025efficientvla}
Yantai Yang, Yuhao Wang, Zichen Wen, Luo Zhongwei, Chang Zou, Zhipeng Zhang, Chuan Wen, and Linfeng Zhang.
\newblock Efficientvla: Training-free acceleration and compression for vision-language-action models.
\newblock \emph{arXiv preprint arXiv:2506.10100}, 2025.

\bibitem[Ye et~al.(2025)Ye, Xie, Zheng, Gao, Wu, Jiang, Li, and Kong]{ye2025dream}
Jiacheng Ye, Zhihui Xie, Lin Zheng, Jiahui Gao, Zirui Wu, Xin Jiang, Zhenguo Li, and Lingpeng Kong.
\newblock Dream 7b: Diffusion large language models.
\newblock \emph{arXiv preprint arXiv:2508.15487}, 2025.

\bibitem[Yi et~al.(2021)Yi, Bai, Tao, Ma, Tian, Wang, Wang, and Fu]{yi2021half}
Jiangyan Yi, Ye~Bai, Jianhua Tao, Haoxin Ma, Zhengkun Tian, Chenglong Wang, Tao Wang, and Ruibo Fu.
\newblock Half-truth: A partially fake audio detection dataset.
\newblock \emph{arXiv preprint arXiv:2104.03617}, 2021.

\bibitem[You et~al.(2025)You, Nie, Zhang, Hu, Zhou, Lu, Wen, and Li]{you2025llada}
Zebin You, Shen Nie, Xiaolu Zhang, Jun Hu, Jun Zhou, Zhiwu Lu, Ji-Rong Wen, and Chongxuan Li.
\newblock Llada-v: Large language diffusion models with visual instruction tuning.
\newblock \emph{arXiv preprint arXiv:2505.16933}, 2025.

\bibitem[Zaheer et~al.(2020)Zaheer, Guruganesh, Dubey, Ainslie, Alberti, Ontanon, Pham, Ravula, Wang, Yang, et~al.]{zaheer2020big}
Manzil Zaheer, Guru Guruganesh, Kumar~Avinava Dubey, Joshua Ainslie, Chris Alberti, Santiago Ontanon, Philip Pham, Anirudh Ravula, Qifan Wang, Li~Yang, et~al.
\newblock Big bird: Transformers for longer sequences.
\newblock \emph{Advances in neural information processing systems}, 33:\penalty0 17283--17297, 2020.

\bibitem[Zhang et~al.(2023{\natexlab{a}})Zhang, Li, Zhang, Zhan, Wang, Zhou, and Qiu]{zhang2023speechgpt}
Dong Zhang, Shimin Li, Xin Zhang, Jun Zhan, Pengyu Wang, Yaqian Zhou, and Xipeng Qiu.
\newblock Speechgpt: Empowering large language models with intrinsic cross-modal conversational abilities.
\newblock \emph{arXiv preprint arXiv:2305.11000}, 2023{\natexlab{a}}.

\bibitem[Zhang et~al.(2024{\natexlab{a}})Zhang, Zhang, Wang, Ouyang, Wen, Li, Chow, He, and Zhang]{zhang2024ocr}
Junyuan Zhang, Qintong Zhang, Bin Wang, Linke Ouyang, Zichen Wen, Ying Li, Ka-Ho Chow, Conghui He, and Wentao Zhang.
\newblock Ocr hinders rag: Evaluating the cascading impact of ocr on retrieval-augmented generation.
\newblock \emph{arXiv preprint arXiv:2412.02592}, 2024{\natexlab{a}}.

\bibitem[Zhang et~al.(2024{\natexlab{b}})Zhang, Fan, Ma, Zheng, Huang, Cheng, Gudovskiy, Okuno, Nakata, Keutzer, et~al.]{zhang2024sparsevlm}
Yuan Zhang, Chun-Kai Fan, Junpeng Ma, Wenzhao Zheng, Tao Huang, Kuan Cheng, Denis Gudovskiy, Tomoyuki Okuno, Yohei Nakata, Kurt Keutzer, et~al.
\newblock Sparsevlm: Visual token sparsification for efficient vision-language model inference.
\newblock \emph{arXiv preprint arXiv:2410.04417}, 2024{\natexlab{b}}.

\bibitem[Zhang et~al.(2023{\natexlab{b}})Zhang, Sheng, Zhou, Chen, Zheng, Cai, Song, Tian, R{\'e}, Barrett, et~al.]{zhang2023h2o}
Zhenyu Zhang, Ying Sheng, Tianyi Zhou, Tianlong Chen, Lianmin Zheng, Ruisi Cai, Zhao Song, Yuandong Tian, Christopher R{\'e}, Clark Barrett, et~al.
\newblock H2o: Heavy-hitter oracle for efficient generative inference of large language models.
\newblock \emph{Advances in Neural Information Processing Systems}, 36:\penalty0 34661--34710, 2023{\natexlab{b}}.

\end{thebibliography}
\bibliographystyle{iclr2026_conference}

\clearpage
\appendix
\startcontents[appendix]
\printcontents[appendix]{ }{0}{\section*{Appendix}}

\section{Additional Results}

\label{sec:additional_results}

\begin{wraptable}{htbp}{0.55\textwidth}
    \centering
    \footnotesize
    \caption{Core statistics of the AudioMarathon.}
    \label{tab:mmau_stats}
    \begin{tabular}{lc}
        \toprule
        \textbf{Statistics} & \textbf{Number} \\
        \midrule
        Total Questions & 6567 \\
        Audio Domains & 10 \\
        Difficulty (Easy:Medium:Hard) & 24\%:61\%:15\% \\
        \midrule
        Speech Content Extraction & 1514 \\
        Automatic Speech Recognition (ASR) & 204 (3.10\%) \\
        Speech Content Reasoning (SCR) & 820 (12.49\%) \\
        Speech Entity Recognition (SER) & 490 (7.46\%) \\
        \midrule
        Audio classification & 1519 \\
        Audio scene classifier (ASC) & 1145 (17.44\%) \\
        Music classifier (MC) & 120 (1.83\%) \\
        Sound event detection (SED) & 254 (3.87\%) \\
        \midrule
        Speaker Recognition & 3530 \\
        Emotion Recognition (ER) & 185 (2.82\%) \\
        Speech Detection (SD) & 776 (11.82\%) \\
        Speaker Age Recognition (SAR) & 959 (14.60\%) \\
        Speaker Gender Recognition (SGR) & 1614 (24.58\%) \\
        \midrule
        Mutiple Choice Questions & 6452 \\
        Transcriptions & 270 \\
        \bottomrule
    \end{tabular}
\end{wraptable}


\textbf{Token Pruning Details.} In multi-modal large language models, the first two layers are regarded as shallow layers,  where attention allocation remains relatively balanced, and output tokens mainly attend to preceding outputs while modality-specific tokens (e.g., vision or audio) are not yet fully integrated into semantic reasoning. Prior analysis on vision tokens has demonstrated that pruning at the second layer is particularly effective: it removes redundant tokens while retaining a compact set of representative ones, thereby preventing redundant information from propagating into deeper layers and significantly reducing computational overhead. Motivated by this observation, we apply the same strategy to audio tokens, pruning them directly at the second layer. This early pruning leverages the redundancy of low-level acoustic representations, which often contain overlapping information, and achieves a favorable balance between efficiency and performance. Compared with pruning at the first layer, where feature representations are still unstable and critical information may be lost, the second layer offers a more appropriate trade-off between stability and efficiency. Conversely, deferring pruning to deeper layers would result in repeated computations on redundant tokens, diminishing overall efficiency. Thus, second-layer pruning of audio tokens can be understood as a low-level information compression mechanism, which eliminates ineffective tokens at an early stage to maximize acceleration in subsequent layers while maintaining robust performance.
\subsection{Random Choice Baseline}
\begin{table*}[h!]
\centering
\small
\caption{Random baseline results on nine subsets of the dataset after 100 random selections. General refers to double-choice questions; Four-choice refers to Four-option choice questions.}
\label{tab:random-baseline}
\begin{tabular}{lcccc}
\toprule
Task & Labels & Type & ACC & Macro F1-score \\
\midrule
MC & 4 & four-choice & 0.2477 & 0.2451 \\
SD  & 2 & general & 0.5200 & 0.4700 \\
SER  & 4 & four-choice & 0.2525 & 0.2518 \\
ASC  & 4 & four-choice & 0.2496 & 0.2493 \\
ER  & 4 & four-choice & 0.2436 & 0.2426 \\
SGR  & 2 & general & 0.5000 & 0.4798 \\
SAR  & 5 & general & 0.1979 & 0.1695 \\
SCR  & 4 & four-choice & 0.2490 & 0.2491 \\
ASC & 10 & general & 0.2467 & 0.2450 \\
\midrule
\multicolumn{3}{c}{Four-choice (6 tasks)} & 0.2490 & 0.2479 \\
\multicolumn{3}{c}{General (3 tasks)} & 0.3999 & 0.3744 \\
\multicolumn{3}{c}{Overall (9 tasks)} & 0.2993 & 0.2901 \\
\bottomrule
\end{tabular}
\end{table*}
The Table~\ref{tab:random-baseline}
illustrates that the F1-scores are lower, directly reflecting the validity of using F1-score as an evaluation metric, highlighting its ability to balance precision and recall.
\subsection{Results of Qwen2.5-Omni-3B on the other six datasets in \benchname{}}
\begin{figure}[h]
    \centering
    \includegraphics[width=1\linewidth]{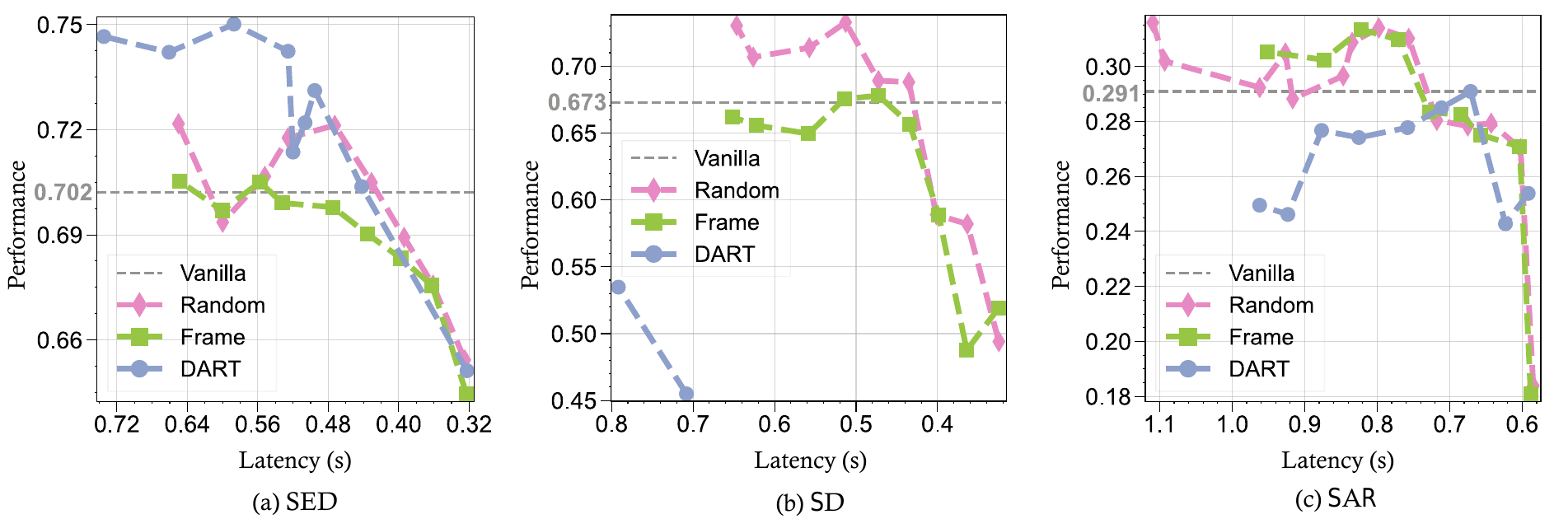}
    \caption{Comparisons of latency and performance trade-off for the Qwen2.5-Omni-3B model on the SED, SD, and SAR dataset}
    \label{fig:latency_bu_1}
\end{figure}

\begin{figure}[h]
    \centering
    \includegraphics[width=1\linewidth]{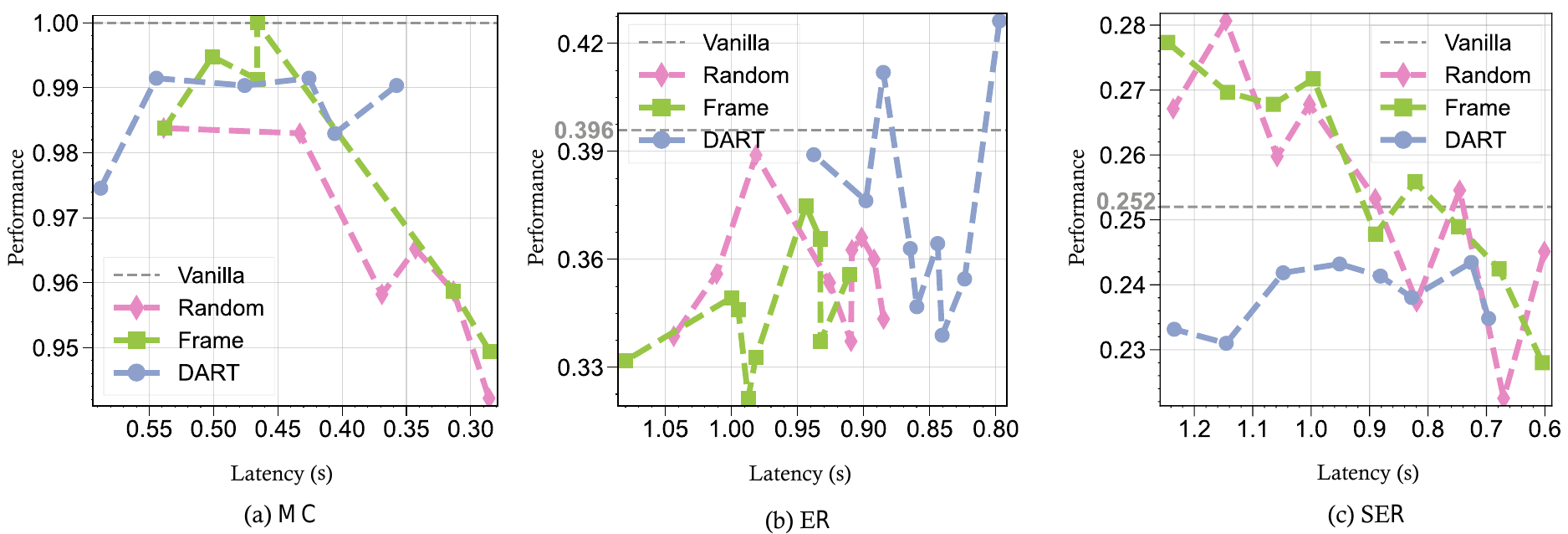}
    \caption{Comparisons of latency and performance trade-off for the Qwen2.5-Omni-3B model on the MC, ER, and SER dataset}
    \label{fig:latency_bu_2}
\end{figure}

While prior results demonstrate the robustness and general superiority of the Frame method on representative MCQs, we further investigate the specific advantages and disadvantages exhibited by certain methods on particular tasks. For instance, on simpler tasks like SED and MC, Frame performs stably, surpassing random methods. However, it underperforms on more challenging tasks, such as SER and SAR. In contrast, Frame maintains robust performance, proving its relative reliability.
\begin{figure}[h]
    \centering
    \includegraphics[width=1\linewidth]{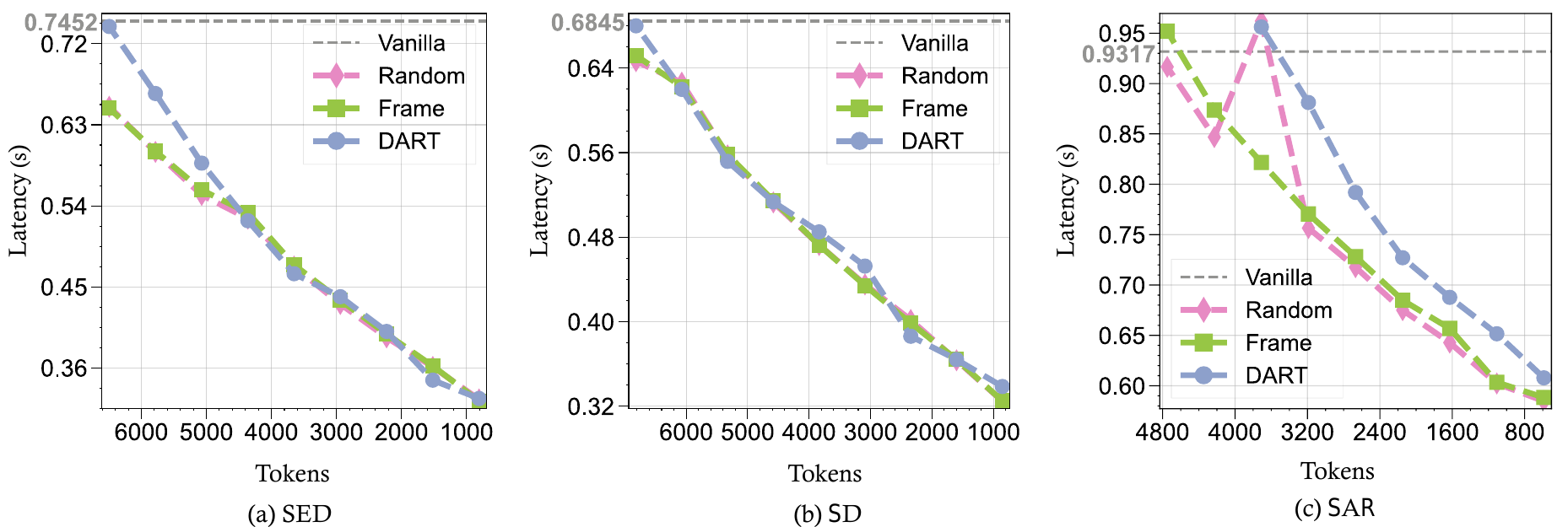}
    \caption{Acceleration effects for the Qwen2.5-Omni-3B model on the SED, SD, and SAR dataset}
    \label{fig:tokens_bu_1}
\end{figure}

\begin{figure}[h]
    \centering
    \includegraphics[width=1\linewidth]{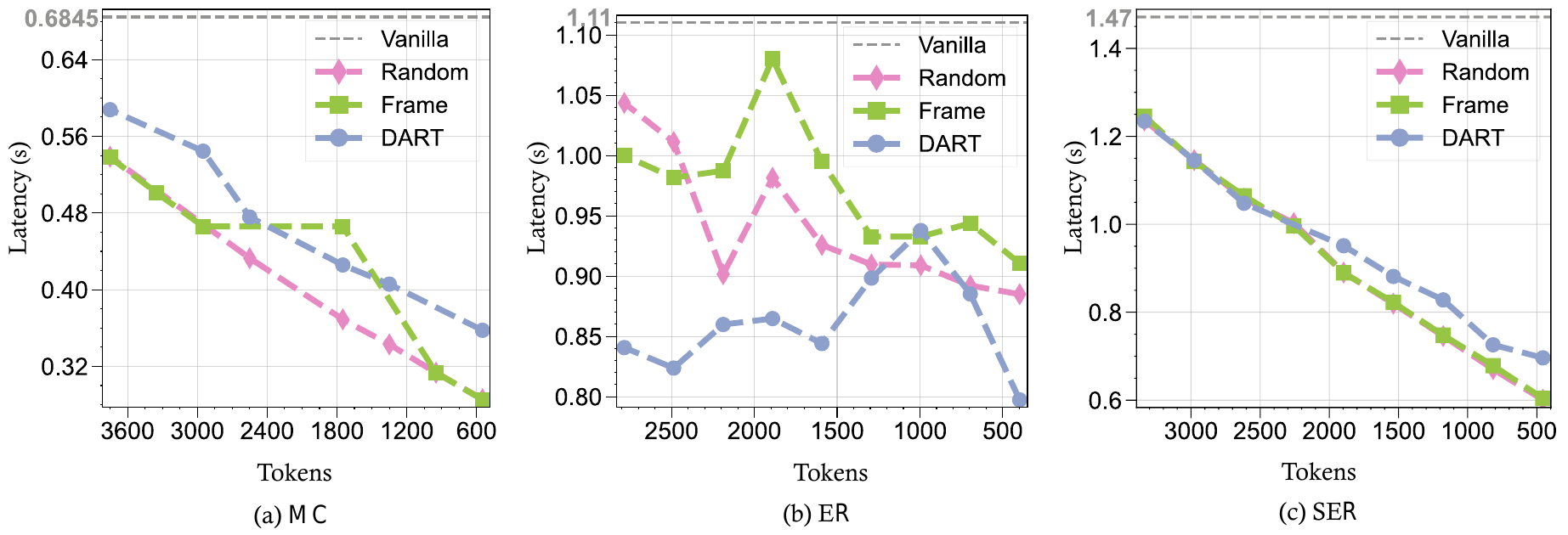}
    \caption{Acceleration effects for the Qwen2.5-Omni-3B model on the MC, ER, and SER dataset}
    \label{fig:tokens_bu_2}
\end{figure}

As shown in Figure~\ref {fig:tokens_bu_1} and Figure~\ref {fig:tokens_bu_2}, different token pruning methods maintain a high degree of consistency in acceleration performance, further illustrating their effectiveness in reducing inference time for long audio MCQ tasks.
\newpage
\definecolor{lightblue}{RGB}{135,206,250} 
\definecolor{lightgreen}{RGB}{144,238,144} 
\definecolor{lightorange}{RGB}{255,218,185} 

\begin{table}[h]
    \caption{Summary of Error Analysis for Phi-4-multimodal on ASR Tasks}
    \label{tab:error_analysis}
    \centering
    \small
    \renewcommand{\arraystretch}{1.2} 
    \begin{tabular}{>{\raggedright\arraybackslash}p{0.25\columnwidth} >{\raggedright\arraybackslash}p{0.7\columnwidth}}
        \toprule
        \textbf{Method} & \textbf{Error Patterns} \\
        \midrule
        \multirow{3}{*}{\textbf{Vanilla}} 
            & \textit{\textcolor{lightblue}{Insertions}}: 'to': 12, 'the': 7, 'a': 4, 'hundred': 3, 'is': 3, 'g.': 3, 'or': 3, 'be': 3, 'with': 3 \\
            & \textit{\textcolor{lightgreen}{Deletions}}: 'the': 13, 'that': 5, 'a': 4, 'percent': 4 \\
            & \textit{\textcolor{lightorange}{Substitutions}}: 'the' $\rightarrow$ 'a': 6, 'in' $\rightarrow$ 'and': 3, 'had' $\rightarrow$ 'has': 3, 'it' $\rightarrow$ 'it's': 3, 'a' $\rightarrow$ 'the': 3 \\
        \midrule
        \multirow{3}{*}{\textbf{FastV Prune 20\%}} 
            & \textit{\textcolor{lightblue}{Insertions}}: 'to': 11, 'the': 8, 'hundred': 4, 'or': 4, 'be': 4, 'and': 3 \\
            & \textit{\textcolor{lightgreen}{Deletions}}: 'the': 451, 'and': 152, 'to': 139, 'of': 126, 'is': 122, 'in': 77, 'a': 76, 'that': 45, 'was': 43, 'are': 40 \\
            & \textit{\textcolor{lightorange}{Substitutions}}: 'the' $\rightarrow$ 'a': 7, 'in' $\rightarrow$ 'and': 3, 'is' $\rightarrow$ 'the': 3, 'a' $\rightarrow$ 'the': 3 \\
        \midrule
        \multirow{3}{*}{\textbf{Dart Prune 20\%}} 
            & \textit{\textcolor{lightblue}{Insertions}}: 'to': 12, 'the': 7, 'a': 5, 'is': 4, 'g.': 3, 'or': 3, 'be': 3, 'with': 3 \\
            & \textit{\textcolor{lightgreen}{Deletions}}: 'the': 52, 'to': 30, 'is': 26, 'are': 13, 'a': 9, 'and': 9, 'that': 8, 'of': 7, 'was': 7, 'in': 6 \\
            & \textit{\textcolor{lightorange}{Substitutions}}: 'the' $\rightarrow$ 'a': 6, 'a' $\rightarrow$ 'the': 4, 'had' $\rightarrow$ 'has': 3 \\
        \bottomrule
    \end{tabular}
\end{table}
\section{Error Analysis}

\label{sec:error_analysis}
In this section, we further compare the token prune results on ASR task. The result of Table~\ref{tab:error_analysis} demonstrates the attention-based selection probably causes the loss of high-frequency words. The model's substantial omission of high-frequency words in the audio transcription task results in inferior performance under the same pruning ratio.

\section{Encoding Granularity of LALMs}
\label{app_sec:Encoding_Granularity}
\begin{table*}[h]
\centering
\scriptsize
\caption{\small Audio processing capacity of Audio Language Models, including maximum supported audio length, maximum number of encoded audio tokens, and token rate (tokens per second).}
\label{tab:audio_models}
\renewcommand{\arraystretch}{1.05}
\sisetup{group-separator={,}}
\setlength{\tabcolsep}{2pt} 
\begin{tabular*}{\textwidth}{@{\extracolsep{\fill}} l c c c}
\toprule
\textbf{Model Name} & \textbf{\begin{tabular}{@{}c@{}}Max Audio\\ Length\end{tabular}} & \textbf{\begin{tabular}{@{}c@{}}Max Encoded\\ Audio Tokens\end{tabular}} & \textbf{\begin{tabular}{@{}c@{}}Token Rate\\ (tokens/s)\end{tabular}} \\
\midrule
Phi-4-multimodal~\citep{abouelenin2025phi} & 30 minutes & {22500} & 12.5 tokens/s \\
Aero-1-Audio~\citep{li2025aero} & 15 minutes & {22500}& 25.0 tokens/s \\
Qwen2-Audio-Instruct~\citep{chu2024qwen2} & 0.5 minutes& {750}& 25.0 tokens/s \\
Qwen2.5-Omni~\citep{xu2025qwen2} & 21 minutes& {32000}& 25.0 tokens/s \\
\bottomrule
\end{tabular*}
\end{table*}
Table~\ref{tab:audio_models} reports the audio encoding granularity of the LALMs. Except for Phi-4-Multimodal, all models produce about 7,500 tokens for a 5-minute clip, even for straightforward tasks such as gender or age classification, which reveals substantial redundancy in current audio encoding.

\section{Dataset Constitute}

\label{sec:dataset_details}

\noindent \textbf{SLUE}~\citep{shon2022slue}.
The Spoken Language Understanding Evaluation (SLUE) benchmark is a suite of tasks designed for evaluating speech models on spoken language understanding. It is derived from the full 960 hours of the LibriSpeech corpus and includes various tasks such as named entity recognition (NER), sentiment analysis, and relation extraction. For \benchname{}, we utilize the sentiment analysis subset, which requires models to comprehend spoken content and infer the underlying sentiment.

\noindent \textbf{RACE}~\citep{lai2017race}.
The Reading Comprehension from Examinations (RACE) dataset is a large-scale collection of reading comprehension questions from English exams for middle and high-school Chinese students. It consists of over 28,000 passages and nearly 100,000 questions written by human experts to evaluate reading comprehension and reasoning skills. In \benchname{}, we use an audio-transcribed version of the RACE dataset, transforming the text-based reasoning challenge into a listening comprehension task that tests a model's ability to process and reason over long spoken narratives. 

\noindent \textbf{LibriSpeech-long}~\citep{park2024long}.
LibriSpeech is a widely used corpus for Automatic Speech Recognition (ASR), containing approximately 1,000 hours of English speech read from public domain audiobooks. The original dataset consists of short audio clips, typically a few seconds long. For \benchname{}, we created LibriSpeech-long by concatenating multiple short clips from the same speaker and chapter to form continuous, long-form audio files, which are used to evaluate the models' long-context ASR performance.

\noindent \textbf{DESED}~\citep{turpault2019sound}.
The Domestic Environment Sound Event Detection (DESED) dataset is designed for the task of sound event detection in real-life domestic environments. The dataset is built using a combination of synthesized and real recordings from AudioSet, focusing on 10 common domestic sound classes (e.g., dog bark, blender, speech). It provides strong annotations with precise event start and end times, making it a challenging benchmark for evaluating the temporal localization capabilities of audio models.

\noindent \textbf{GTZAN}~\citep{tzanetakis2002musical}.
GTZAN Genre Collection is one of the most widely used datasets for music genre classification. It consists of 1,000 audio tracks, each 30 seconds long, distributed evenly across 10 distinct music genres (e.g., Blues, Classical, Hip-Hop, Jazz, Rock). Each genre is represented by 100 clips. Despite some known issues with label consistency in a small fraction of the data, it remains a standard benchmark for evaluating music information retrieval.

\noindent \textbf{TAU Urban Acoustic Scenes}~\citep{heittola2019tau}.
TAU Urban Acoustic Scenes dataset is a collection of recordings from various acoustic scenes for the task of acoustic scene classification. The 2019 version, which we reference, contains over 40 hours of audio recorded in 10 different European cities. The data is provided as 10-second segments extracted from longer original recordings, capturing diverse urban environments such as airports, public parks, and metro stations. In our benchmark, we utilize these longer source recordings to test scene classification in extended audio contexts.

\noindent \textbf{HAD}~\citep{yi2021half}.
The Hallym Aging Diacrisis (HAD) dataset is a Korean speech corpus designed for the study of age-related voice characteristics and the diagnosis of pathological voices in the elderly. It contains speech samples from different age groups, including young adults and elderly individuals, performing various speech tasks like reading passages and sustained vowel phonations. The dataset is annotated with speaker age and health status, making it suitable for tasks related to age estimation and vocal health detection.

\noindent \textbf{VESUS}~\citep{sager2019vesus}.
The Voice Evaluation for Specific UtteranceS (VESUS) dataset is a corpus for assessing voice pathologies. It contains recordings from speakers with various voice disorders as well as healthy controls. Speakers were recorded producing specific utterances, such as sustained vowels and standard sentences, which are designed to highlight vocal impairments. The dataset is annotated by expert clinicians with labels for overall voice quality and specific perceptual ratings (e.g., roughness, breathiness, strain), serving as a benchmark for automated voice quality assessment systems.

\noindent \textbf{Vox\_Age \& Vox\_Gender}~\citep{hechmi2021voxceleb}.
These tasks are derived from the large-scale VoxCeleb speaker recognition dataset~\citep{nagrani2017voxceleb}. VoxCeleb consists of hundreds of thousands of "in-the-wild" speech clips extracted from celebrity interview videos on YouTube. While the primary purpose of VoxCeleb is speaker identification and verification, the metadata associated with each celebrity allows for the creation of auxiliary tasks. For \benchname{}, we use this data to evaluate speaker characteristic identification, specifically age estimation (VoxAge) and gender classification (VoxGender) from long, unconstrained speech segments.

\section{Model Details}
\label{sec:model_details}
\textbf{Phi-4-Multimodal}~\citep{abouelenin2025phi}.
This model is extended from Phi-4-Mini and integrates three input modalities: text, vision, and speech/audio. Its key innovation lies in the use of the ``Mixture-of-LoRAs'' technique: while keeping the base language model completely frozen, it introduces modality-specific LoRA adapters and a routing mechanism to enable flexible multimodal reasoning (e.g., vision + language, vision + speech, speech-only) without interference across modalities.  

\textbf{Qwen2.5-Omni}~\citep{xu2025qwen2}.
Developed by the Qwen Team, this is an end-to-end multimodal model capable of perceiving multiple modalities, including text, image, audio, and video, and supporting streaming generation of both text and natural speech responses. Its main innovations include: the introduction of TMRoPE (temporally aligned multimodal rotary position embedding) for audio-video timestamp synchronization; the Thinker–Talker architecture, where the Thinker is responsible for text generation and the Talker generates audio tokens based on the hidden states of the Thinker, thereby avoiding interference between text and speech generation; and the use of block-level processing and sliding-window DiT mechanisms to reduce streaming latency.  

\textbf{Audio-Flamingo-2 (AF2)}~\citep{ghosh2025audio}.
This model is an audio language model (ALM) with long audio understanding ability (30 seconds to 5 minutes) and expert-level reasoning capabilities. Its core innovations include: the AF-CLAP audio encoder, trained with an improved contrastive loss on over 8 million audio–text pairs; the AudioSkills dataset, which consists of 4.2 million question–answer pairs covering seven categories of reasoning skills; and a three-stage curriculum training strategy including pretraining, fine-tuning, and long-audio fine-tuning.  

\textbf{Audio-Flamingo-3 (AF3)}~\citep{goel2025audio}.
Jointly developed by NVIDIA and the University of Maryland, this is a leading fully open-source LALM. Its main innovations include: the AF-Whisper unified audio encoder, which enables joint representation learning of speech, environmental sounds, and music; support for on-demand reasoning (e.g., chain-of-thought reasoning), multi-turn multi-audio dialogue, long audio understanding up to 10 minutes, and speech-to-speech interaction.  

\textbf{Baichuan-Omni-1.5}~\citep{li2025baichuan}.
Developed by Baichuan Inc., this is a full-modality model capable of understanding text, image, audio, and video, as well as supporting end-to-end audio generation. Its main strengths include: a data processing pipeline that constructs and cleans approximately 500B high-quality multimodal data; the Baichuan-Audio-Tokenizer, designed to capture both semantic and acoustic features (implemented with an 8-layer RVQ structure and a 12.5 Hz frame rate); and a multi-stage training strategy consisting of image–text pretraining, image–audio–text joint pretraining, full-modality pretraining, and multimodal supervised fine-tuning.  

\textbf{Gemma-3n}~\citep{gemma_3n_2025}.
The Gemma 3n models are optimized for efficient execution on low-resource devices. They support multimodal input (text, image, video, audio) and generate high-quality text outputs. The series provides open weights for both pre-trained and instruction-tuned variants, and covers more than 140 natural languages. The Gemma 3n models employ selective parameter activation technology, which reduces resource requirements and allows the models to operate effectively at sizes of 2B or 4B parameters, although the total number of parameters is larger.  

\textbf{Aero-1-Audio}~\citep{li2025aero}. 
Aero-1-Audio is a compact audio model developed by LMMs-Lab as part of the Aero-1 series, the first generation of lightweight multimodal systems. Built upon the Qwen-2.5-1.5B language model, it achieves strong performance across speech recognition, audio understanding, and instruction-following benchmarks while remaining parameter-efficient. Trained within one day on 16 H100 GPUs with 50k hours of curated data, Aero demonstrates that efficient training is possible with high-quality samples. It further supports continuous audio inputs up to 15 minutes, a challenging setting for most existing audio models.  

\textbf{GPT-4o}~\citep{hurst2024gpt}.
GPT-4o, released by OpenAI in August 2024, is an autoregressive universal model supporting arbitrary combinations of text, audio, image, and video as inputs, and text, audio, and image as outputs. All modalities are processed by a single end-to-end trained neural network, enabling seamless multimodal integration and efficient inference across diverse tasks.  

\textbf{Gemini-2.0-Flash-Lite}~\citep{comanici2025gemini}.
Gemini-2.0-Flash-Lite, introduced by Google in April 2025, is the most cost-efficient member of the Gemini 2.0 family. It adopts a sparse Mixture-of-Experts Transformer architecture and leverages Trillium TPUs for training and inference. The model supports text, image, audio, and video inputs with a context window of 1,048,576 tokens, and produces text outputs of up to 8,192 tokens. Its design emphasizes scalability and latency efficiency for high-volume multimodal applications.  

\textbf{Gemini-2.0-Flash}~\citep{comanici2025gemini}.
Gemini-2.0-Flash is a natively multimodal model designed to power next-generation agentic systems. Compared with Gemini 1.5 Flash, it offers higher quality while maintaining comparable inference speed. It accepts text, image, audio, and video inputs with a 1,048,576-token context window and outputs text up to 8,192 tokens, with experimental image generation capabilities. Its architecture refines the sparse Mixture-of-Experts Transformer design with improved stability and optimization efficiency.  

\textbf{Gemini-2.5-Flash}~\citep{comanici2025gemini}.
Gemini-2.5-Flash is Google’s first hybrid reasoning model, allowing developers to toggle reasoning on or off and allocate reasoning budgets for a trade-off between quality, cost, and latency. It supports text, image, audio, and video inputs with a 1M-token context window and generates text outputs up to 64K tokens. Based on a sparse Mixture-of-Experts Transformer with native multimodal support, it significantly outperforms Gemini-1.5-Pro on reasoning and multimodal benchmarks.  

\textbf{Gemini-2.5-Flash-Lite}~\citep{comanici2025gemini}.
Gemini-2.5-Flash-Lite extends the hybrid reasoning family with a cost-efficient design optimized for latency-sensitive tasks such as translation and classification. It provides improvements over Gemini-2.0-Flash-Lite in coding, mathematics, science, and reasoning, while supporting text, image, audio, and video inputs with a 1M-token context window and generating text outputs up to 64K tokens. Its sparse Mixture-of-Experts Transformer architecture balances efficiency with strong performance in large-scale multi-modal applications.


\raggedbottom 

\section{Prompt}

Here we present the prompt templates used for various tasks in our \benchname{}.

\begin{figure}[htp]
    \centering
    \includegraphics[width=\textwidth]{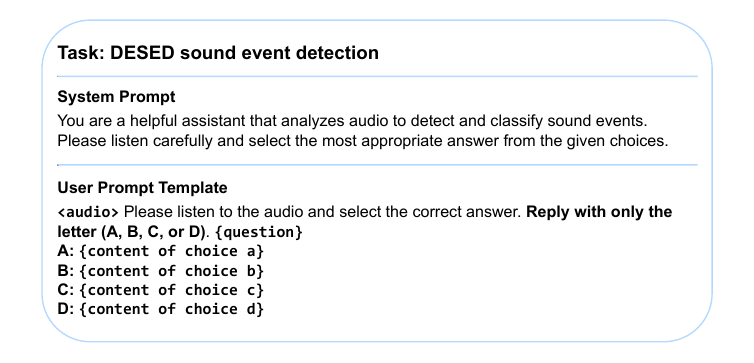}
    \caption{Prompt template for the SED task.}
    \label{fig:desed}
\end{figure}

\begin{figure}[htp]
    \centering
    \includegraphics[width=\textwidth]{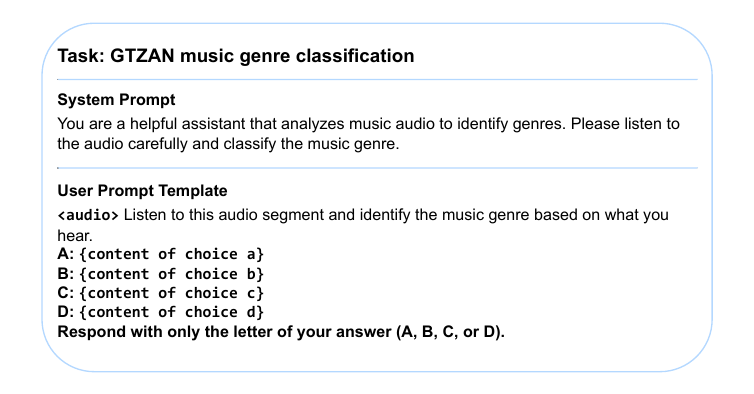}
    \caption{Prompt template for the MC task.}
    \label{fig:gtzan}
\end{figure}

\begin{figure}[htp]
    \centering
    \includegraphics[width=\textwidth]{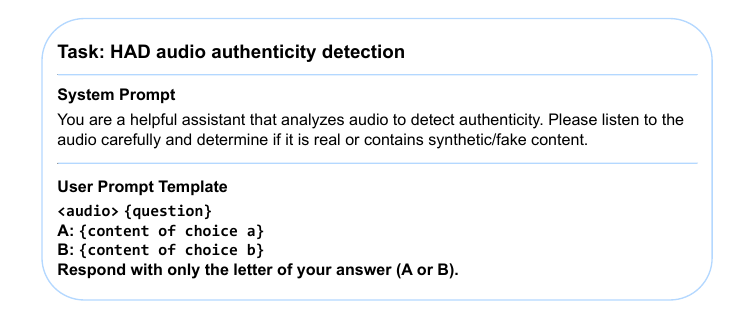}
    \caption{Prompt template for the SD task.}
    \label{fig:had}
\end{figure}

\begin{figure}[htp]
    \centering
    \includegraphics[width=\textwidth]{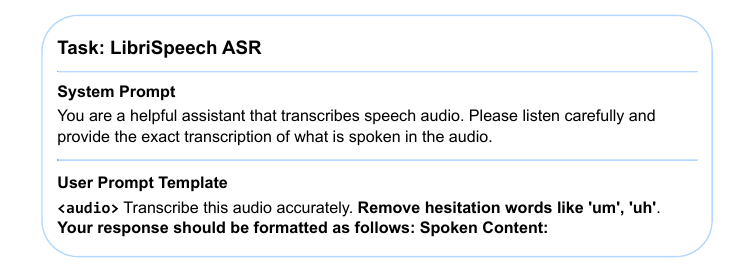}
    \caption{Prompt template for the ASR task.}
    \label{fig:librispeech}
\end{figure}

\begin{figure}[htp]
    \centering
    \includegraphics[width=\textwidth]{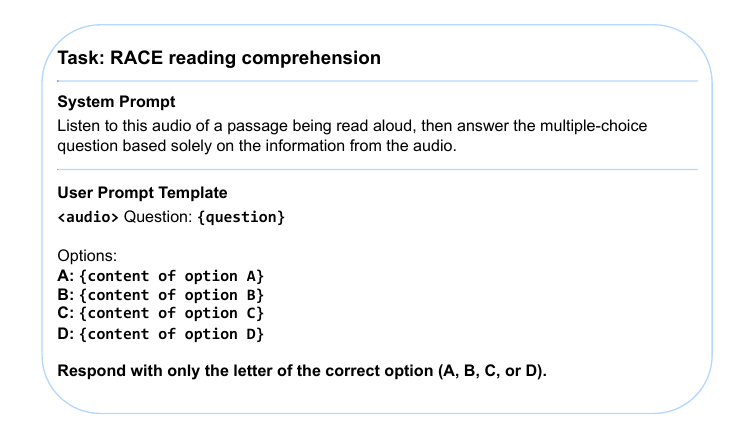}
    \caption{Prompt template for the SCR task.}
    \label{fig:race}
\end{figure}

\begin{figure}[htp]
    \centering
    \includegraphics[width=\textwidth]{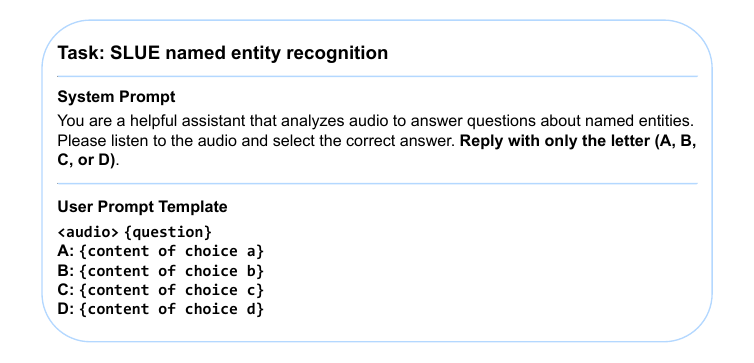}
    \caption{Prompt template for the SER task.}
    \label{fig:slue}
\end{figure}

\begin{figure}[htp]
    \centering
    \includegraphics[width=\textwidth]{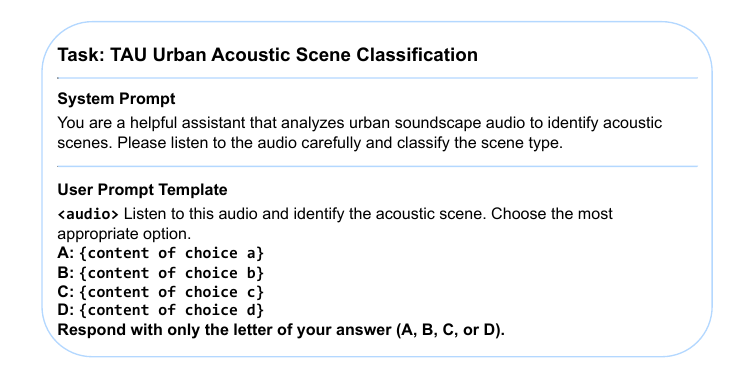}
    \caption{Prompt template for the ASC task.}
    \label{fig:tau}
\end{figure}
\begin{figure}[htp]
    \centering
    \includegraphics[width=\textwidth]{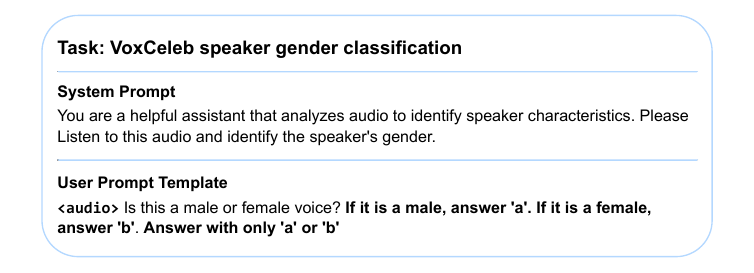}
    \caption{Prompt template for the SGR task.}
    \label{fig:vox-gender}
\end{figure}
\begin{figure}[htp]
    \centering
    \includegraphics[width=\textwidth]{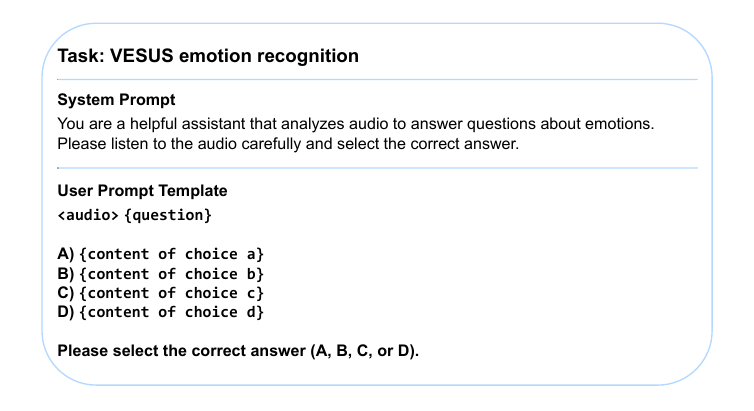}
    \caption{Prompt template for the ER task.}
    \label{fig:vesus}
\end{figure}

\begin{figure}[htp]
    \centering
    \includegraphics[width=\textwidth]{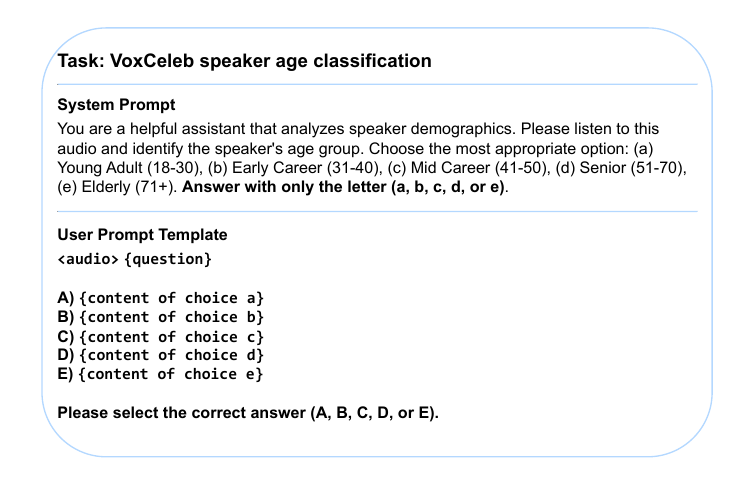}
    \caption{Prompt template for the SAR task.}
    \label{fig:vox-age}
\end{figure}

\clearpage


\section{Limitations and Future Work}

\subsection{Limitations}
Our benchmark reflects practical choices in data sources and task design. Some datasets have license limits that restrict redistribution. The benchmark focuses on English and may not reflect cross-language behavior. We rely on automatic pipelines for audio concatenation and option generation, which can introduce bias if the source data has bias. While we test multiple long audio tasks, some domains and tasks are still underrepresented. Our evaluation covers common metrics but does not fully capture human preference or safety risks. Finally, we focus on inference efficiency methods at test time and do not include training time efficiency or energy use.

\subsection{Future Work}
We plan to run systematic hyperparameter searches at key encoder and decoder layers to measure sensitivity and find settings that preserve temporal detail while improving efficiency. We will evaluate more compression and acceleration methods, including stronger token selection methods and better cache policies, and test transfer from text and vision methods to audio. We will add more tasks and languages, broaden source datasets, and release tools for reproducible data building and evaluation. We also plan to study human evaluation for long audio tasks and extend metrics that measure temporal continuity and memory. Finally, we will report energy and cost to give a fuller view of efficiency.
In this work, we observe that existing autoregressive audio LLMs still face a dual challenge of accuracy and efficiency when processing long-form audio, as exemplified by ASR. Exploring architectures that depart from the autoregressive paradigm may offer a promising path forward. In particular, the recent surge of diffusion LLMs~\citep{nie2025large, you2025llada, ye2025dream, wen2025devil,jin2025thinking}, which support parallel decoding, suggests a compelling alternative. Leveraging diffusion LLMs for long-audio understanding and generation could unlock substantial gains in both fidelity and computational throughput.

\section{Use of LLMs}
In this study, we utilized large language models (LLMs) to perform grammar checking and to polish certain sentences for improved clarity and fluency, without altering the original meaning of the text. Auxiliary AI coding tools are used for debugging and analyzing code errors, as well as assisting in code implementation, with the main code being constructed and carefully reviewed by humans.

\end{document}